\documentclass[aps,prb,twocolumn,amsmath,amssymb,nofootinbib,superscriptaddress,floatfix]{revtex4-1}
\usepackage{amssymb}
\usepackage{mathrsfs}
\usepackage[hypertex]{hyperref}
\usepackage{color}
\usepackage{graphicx}
\usepackage{srcltx}

\def\AA{{\cal A}}
\def\LL{{\cal L}}

\def\MM{{\cal M}}

\def\PP{\partial}
\def\SS{{\cal S}}

\begin{document}

\title{
Topological Phases on Non-orientable Surfaces: Twisting by Parity Symmetry
}

\author{AtMa P.O. Chan}
\affiliation{
Institute for Condensed Matter Theory and
Department of Physics, University of Illinois at Urbana-Champaign, 1110 West Green St, Urbana IL 61801}

\author{Jeffrey C.Y. Teo}
%\email{jteo@virginia.edu}
\affiliation{Department of Physics, University of Virginia, Charlottesville, VA 22904, USA}

\author{Shinsei Ryu}
\affiliation{
Institute for Condensed Matter Theory and
Department of Physics, University of Illinois at Urbana-Champaign, 1110 West Green St, Urbana IL 61801}

%\date{\today}

\begin{abstract}
We discuss (2+1)D topological phases on non-orientable spatial surfaces, such as M\"obius strip, real projective plane and Klein bottle, etc., which are obtained by twisting the parent topological phases by their underlying parity symmetries through introducing parity defects. We construct the ground states on arbitrary non-orientable closed manifolds and calculate the ground state degeneracy. Such degeneracy is shown to be robust against continuous deformation of the underlying manifold. We also study the action of the mapping class group on the multiplet of ground states on the Klein bottle. The physical properties of the topological states on non-orientable surfaces are deeply related to the parity symmetric anyons which do not have a notion of orientation in their statistics. For example, the number of ground states on the real projective plane equals the root of the number of distinguishable parity symmetric anyons, while the ground state degeneracy on the Klein bottle equals the total number of parity symmetric anyons; In deforming the Klein bottle, the Dehn twist encodes the topological spins whereas the Y-homeomorphism tells the particle-hole relation of the parity symmetric anyons.

\end{abstract}

\maketitle
\tableofcontents

\section{Introduction}

One of the most striking discoveries in condensed matter physics in the last few decades has been topological states of matter \cite{QFTMBS, FTOCMS, ANYONSKIT, TOPOQC}, in which the characterization of order goes beyond the paradigm of Landau's symmetry breaking theory. Topological states are gapped quantum many-body systems in which the topological order is characterized by their fractional excitations called anyons \cite{ANYONSKIT, TOPOQC, INTERFER}. These emergent anyons have non-trivial braiding statistics, i.e., dragging an anyon once around another causes a unitary transformation on the quantum state of the system. In particular, the braiding of anyons gives rise to a phase factor in abelian topological states. With a non-trivial anyon content $\AA$, topologically ordered states exhibit long-range entanglement \cite{TOPOENEN} and a robust ground state degeneracy (GSD) on torus \cite{GSD1, GSD2}.

If the microscopic Hamiltonian underlying a topologically ordered state possesses a symmetry dictated by a symmetry group $G$, such symmetry can interplay with the topological order described by $\AA$, giving rise to different symmetry enriched topological (SET) states \cite{TIs, TIsSCs, AVSPT, ABELIANSET, SETCLASS, TWISTLIQUID, SYMDEFGAUGE}. If a symmetry $\MM$ in $G$ is on-site, it permutes the anyon labels in such a way that the braiding statistics is preserved. However, if a symmetry $\MM$ in $G$ involves parity, the anyon relabeling changes the braiding statistics by a conjugation. In addition, such parity symmetry also changes the anyon position. We take the parity to be acting in $x$-direction in this paper; so the $x$-coordinate of any anyon is flipped under the parity symmetry.

One way in which a symmetry can intertwine with the topological properties of a state is to twist the theory by the symmetry through introducing defects \cite{TWISTSDEF, DEFECT, LATDIS, GENON, CONG, BFDEFECT, ANYSYM, TWISTLIQUID, SYMDEFGAUGE}. Associated with each symmetry $\MM$ in $G$, we can introduce a point-like twist defect from which a branch cut emanates. Any anyon passing across the branch cut gets transformed or twisted by the symmetry action $\mathcal{M}$. Unlike the anyons which are finite-energy excitations, the defects can only be extrinsically imposed by suitably modifying the microscopic Hamiltonian. Moreover, the defects can braid with projective non-abelian statistics and trap exotic zero modes. The study of physical properties of the defects helps us to gain understanding about the topological state itself in the presence of the symmetry group $G$. For example, the different possible ways by which defects can fuse and associate with each other classify the SET states under $G$ \cite{TWISTLIQUID, SYMDEFGAUGE}.

Twisting by parity opens a window for us to study parity symmetric states defined on non-orientable surfaces. Similar to on-site symmetries, parity can also be used to define defects and to twist topological phases. Though the construction of defects for the parity is similar to that for the on-site symmetries, the parity defects differ from the on-site defects in that they affect the anyon position. In addition to the anyon relabeling, the parity defects also flip the $x$-coordinate for any anyon passing through the branch cut. In other words, such branch cut acts like flipping over the surface spatially. Consequently, by suitably adding the parity defects, we can turn the underlying surface non-orientable. While topological actions generally fail to be defined on non-orientable surfaces \cite{JMLEE, Chen2014}, introducing parity branch cuts allow us to discuss topological phases on non-orientable surfaces.

A key question which we aim to answer in this paper is what we can learn by putting a parity symmetric topological state on non-orientable surfaces; We may gain understanding to the state itself by examing how it responds to different topologies \cite{GSD1, GSD2, BoundaryD}. This is partly motivated by previous studies on the edge states supported by (2+1)D parity-protected topological phases \cite{SPTANOMALY3, SPTANOMALY4}. In these studies, the
edge theory, which is a conformal field theory (CFT), is put on a non-orientable {\it spacetime} \cite{Fuchs2004, SPTANOMALY3, SPTANOMALY4} by parity twist, defining an orientifold CFT. Putting the edge theory on a non-orientable spacetime allows us to infer the triviality or non-triviality of the corresponding bulk state \cite{SPTANOMALY3, SPTANOMALY4}: the orientifold edge theory suffers from quantum anomalies when the bulk is non-trivial whereas it is anomaly-free when the bulk is trivial. Here we aim to figure out what we can learn about the bulk state by putting it on non-orientable spatial surfaces..

\subsection{Overview}

In this paper, we focus on $(2+1)$D parity symmetric abelian topological states, though the discussion can be extended to non-abelian states. Our goal here is to study the physical properties which show up when a state is put on non-orientable surfaces. In our setup, the non-orientable spatial surfaces are generated by introducing parity defects. We first formulate these parity defects in terms of microscopic degrees of freedom so as to have a microscopic understanding of the non-orientable surfaces. Then we consider putting a state on a ribbon geometry and subsequently twist the theory by parity symmetry through introducing a parity branch cut. In the parity-twisted theory, we show that the orientation of any loop gets flipped when the loop is carried once around the ribbon, meaning that such twisted ribbon is actually a M\"obius strip $\mathbb{M}$. Once the state is put on a M\"obius strip $\mathbb{M}$, we find that the anyons which are invariant under the parity symmetry show up as the relevant anyons under the twist. Interestingly, these parity symmetric anyons exhibit a loss of orientation, in the sense that their counter-clockwise statistics is identified with their clockwise statistics.

To proceed, we consider putting the state on non-orientable closed surfaces. By compactifying the M\"obius strip $\mathbb{M}$, which emerges as a result of twisting by the parity symmetry, we get the real projective plane $\mathbb{P}$. While the torus $\mathbb{T}$ generates the orientable closed surfaces, similarly, the real projective plane $\mathbb{P}$ generates the non-orientable closed surfaces. For example, gluing two real projective planes together gives the Klein bottle $\mathbb{K}$ whereas gluing three real projective planes together gives the Dyck's surface. Hence we can construct any arbitrary non-orientable closed manifold with the real projective plane $\mathbb{P}$. It is well known that topological states have topology-dependent GSD on orientable closed surfaces \cite{GSD1, GSD2}. Here, we construct the ground states and calculate the GSD on arbitrary non-orientable closed manifolds. The main result for the GSD is summarized in Eqs.\ (\ref{GSDoddP}) and (\ref{GSDevenP}). The GSD depends on both the topology of the surface and the parity symmetry we used in constructing the surface. In particular, the GSD on the real projective plane $\mathbb{P}$ equals the root of the number of {\it distinguishable} parity symmetric anyons (For the definition of distinguishable symmetric anyons, see Sec.\ \ref{S2C}), whereas the GSD on the Klein bottle $\mathbb{K}$ equals the total number of parity symmetric anyons. Besides, we also show that the calculated GSD is consistent with the Dyck's theorem, and hence is robust against continuous deformation of the non-orientable surface.

Implications of our main result on the GSD on non-orientable closed manifolds can be well described by comparing the following two topological states of matter: the double-semion state and the $\mathbb{Z}_2$ toric code, where the parity symmetry in each the two states is unique. These two states share the same GSD on the torus, $\mathrm{GSD}(\mathbb{T})=4$, and the same total quantum dimension, and hence cannot be distinguished by the topological entanglement entropy. When put on non-orientable surfaces, however, their GSDs now distinguish these two states. For example, $\mathrm{GSD}(\mathbb{P})=1$ for the double-semion state, while $\mathrm{GSD}(\mathbb{P})=2$ for the toric code. Such difference owes to the fact that the number of distinguishable parity symmetric anyons in the two states are different. Besides, $\mathrm{GSD}(\mathbb{K})=2$ for the double-semion state, whereas $\mathrm{GSD}(\mathbb{K})=4$ for the toric code. Their GSDs are different on the Klein bottle $\mathbb{K}$ because they have different number of parity symmetric anyons. Actually, their GSDs are different on any non-orientable closed manifold.

Finally, we study the physical action of the symmetry group of the manifold called the mapping class group (MCG). On the torus $\mathbb{T}$, the MCG is generated by the modular S-transformation and the Dehn twist. By studying the action of the symmetries on the ground states, one can obtain a matrix representation of the MCG. In that case, the S-transformation tells the braiding statistics of the anyons whereas the Dehn twist tells the exchange statistics of the anyons. Motivated by the calculations on the torus $\mathbb{T}$,
we do the same thing on non-orientable closed manifolds. Knowing that the real projective plane $\mathbb{P}$ has trivial MCG, we consider the first non-trivial case which is the Klein bottle $\mathbb{K}$. The MCG of the Klein bottle $\mathbb{K}$ is generated by a Dehn twist and a Y-homeomorphism. Our result for the matrix representation of the MCG is shown in Eq.\ (\ref{MRT}) and (\ref{MRY}). From the matrix representation of the MCG, we find that the Dehn twist tells the exchange statistics of the parity symmetric anyons and the Y-homeomorphism acts like particle-hole conjugation.

\subsection{Organization of the paper}

The content of the paper is organized as follows.
In Sec.\ \ref{S2}, we review the K-matrix theory which serves as a
mathematical framework in describing $(2+1)$D abelian topological states and we incorporate symmetries into the framework. Also, we introduce the notion of symmetric anyons.
In Sec.\ \ref{S3}, we define the defect branch cut associated to each symmetry operation $\MM$. In addition, we show that any anyon getting across the branch cut is twisted by the symmetry $\MM$ through examining the boundary condition imposed by the branch cut.
In Sec.\ \ref{S5}, We construct the M\"obius strip $\mathbb{M}$ by suitably adding a parity branch cut on a ribbon, showing that the parity symmetric anyons become the only relevant anyons as a result of twisting by parity symmetry. Then we compactify it to get the real projective plane $\mathbb{P}$ and hence all the non-orientable closed surfaces. In each of the cases, we construct the ground states and calculate the GSD. Then we show that the GSD is robust against continuous deformation of the manifold.
In Sec.\ \ref{S6}, we study the action of the MCG on the ground states of the Klein bottle $\mathbb{K}$ and obtain a matrix representation for the MCG.
Sec.\ \ref{S7} provides concrete examples for the theoretical discussions made. We also illustrate that, in certain scenario, a pair of parity defects can be viewed as a deformed genon \cite{TWISTSDEF, LATDIS}.
Appendix\ \ref{S4} provides a detailed discussion for the parity symmetric abelian states. We show a structural property for the set of parity symmetry anyons which turns out to be very useful in the derivations.

\section{Abelian Topological States}\label{S2}

We start by introducing the field theoretical framework, the K-matrix theory, for abelian topological states in $(2+1)$D. Then we proceed to incorporate symmetries in the K-matrix theory. At the end of this section, we introduce the concept of symmetric anyons.

\subsection{K-matrix Theory}

An abelian topological state is a state which host anyons obeying abelian statistics. Any abelian state is described by an abelian Chern-Simons theory known as the K-matrix theory, and is specified by an invertible symmetric integer matrix $K$ \cite{QFTMBS}. Consider an abelian topological state characterized by $K$ with $\mathrm{Dim}\, K=N$, which is described by the following (2+1)D Lagrangian:
\begin{align}\label{bulklagrangian}
\LL_{\textrm{Bulk}}=\frac{1}{4\pi}\varepsilon^{\mu\nu\rho}a_{\mu}^TK\PP_{\nu}a_{\rho},
\end{align}
where $a_\mu$ is an $N$-component compact $U(1)$ gauge field. It describes a phase of matter which is gapped in the bulk. In the presence of a boundary, such abelian Chern-Simons theory induces a bosonic Lagrangian at the edge,
\begin{align}\label{edgelagrangian}
\LL_{\textrm{Edge}}=\frac{1}{4\pi}\PP_x\phi^TK\PP_t\phi,
\end{align}
where $\phi$ is an $N$-component bosonic field compactified by $2\pi\mathbb{Z}^N$. In Eq.\ref{edgelagrangian}, we have omitted the velocity term which comes up from the microscopic physics at the edge. Such lagrangian describes $N$ branches of gapless edge modes. Each positive eigenvalue of $K$ corresponds to a left-moving branch and each negative eigenvalue corresponds to a right-moving branch.

In the topological state $K$, any quasi-particle excitation is labeled by an integer vector $l\in\mathbb{Z}^N$, in which the quasi-particle fusion corresponds to vector addition. Local particles are the constituent particles in the microscopic Hamiltonian and they occupy the sublattice $K\mathbb{Z}^N$. Identifying any two excitations differed by a local particle, we define the anyon lattice as
\begin{align}
\AA=\mathbb{Z}^N/K\mathbb{Z}^N.
\end{align}
The anyon lattice $\AA$ forms an abelian group with order $|$det$K|$ under the fusion rule.
In the bulk, the Wilson operator $W^l_\gamma=\exp{ il^T\int_\gamma a_\mu dx^\mu}$
carries anyon $l$ along a path $\gamma$. If $\gamma$ possesses open ends in the bulk, the Wilson operator generates a gapped particle-hole excitation. On the other hand, there is no energy cost if $\gamma$ forms a loop, or ends at the parity boundary. At the spatial boundary, the field operator $\psi^l(x)=e^{il^T\phi(x)}$ annihilates the anyon $l$ at the position $x$ of the edge. By braiding and exchanging the anyons, we get the mutual- and self-statistics,
\begin{align}
\theta_{ll'}&=2\pi l^TK^{-1}l'~~\textrm{mod}~2\pi,\\
\theta_{l}&=\pi l^TK^{-1}l~~\textrm{mod}~2\pi/\pi.\label{exchange}
\end{align}
Local particles are mutually bosonic in the sense that $\theta_{ll'}=0~~\textrm{mod}~2\pi$.
In addition, any local-particle is either self-bosonic with $\theta_{l}=0~\textrm{mod}~2\pi$ or self-fermionic with $\theta_{l}=0~\textrm{mod}~\pi$. So in Eq.\ (\ref{exchange}), the self-statistics of anyons is defined up to $2\pi$ for bosonic systems and it is determined only up to $\pi$ if there exists fermionic local particles. Let $\mathcal{D}=|\textrm{det}K|^{1/2}$ is the total quantum dimension of the state, the $S$ and $T$ matrices are defined by
\begin{align}\label{ST}
S_{ll'}=e^{i\theta_{ll'}}/\mathcal{D}
~~~~\mbox{and}~~~~
T_{ll'}=\delta_{ll'}e^{i\theta_{l}},
\end{align}
where the T matrix is defined up to a sign if there exists fermionic local particles. Any abelian state $K$ is modular in the sense that $S$ is unitary. If a state is modular, the $S$ and $T$ matrices together with the charge conjugation matrix $C_{ll'}=\delta_{\bar{l}l'}$, where $\bar{l}=-l$ is the particle conjugation of the anyon $l$, satisfy the modular relations \cite{ANYONSKIT}.
In the K-matrix formalism, all the statistical properties of the anyons are encoded in the matrix $K$.

%The topological state can be coupled to a $U(1)$ gauge field $A_{\mu}$. In such case, the state $K$ is equipped with an additional integer vector $t$ which tells how the state is coupled to the gauge field. If the gauge field is electromagnetic, the vector $t$ is called the charge vector. If the gauge field is the axial gauge field, the vector $t$ is called the spin vector. With the vector $t$, each anyon $l$ is associated with a physical quantity $q^l_{t}=t^TK^{-1}l~~\textrm{mod}~1$,
%\begin{align}
%q^l_{t}=t^TK^{-1}l~~\textrm{mod}~1,
%\end{align}
%where we assume the smallest charge carried by local particles is one. Since $q^l_{t}$ is linear in $l$ and we have $q^{l+l'}_{t}=q^{l}_{t}+q^{l'}_{t}$. Under the external gauge field $A_{\mu}$, the topological state is insulating in the sense that the longitudinal conductance is zero whereas the Hall conductance is quantized to $\sigma^{xy}_t=\frac{1}{2\pi}t^TK^{-1}t$.

\subsection{Formulation of Symmetries}

Here we discuss the formulation of symmetry within the framework of K-matrix theory \cite{AVSPT, MLSPT, ANYSYM, CPTSYM}. Any transformation $\MM$ is defined by specifying its action on the local particle operators. Such action can be rewritten as a transformation on the bulk and edge field variables in the K-matrix theory. Consider a state $K$ defined on upper x-y plane with an edge along the x-axis. Let $\tilde{v}^\mu=(v^0,~sv^1,~v^2)$ for any vector $v^\mu$. In general,
\begin{align}\label{sym}
\MM a_\mu(x^\nu)\MM^{-1}&=U^T\tilde{a}_\mu(\tilde{x}^\nu),\\
\MM\phi(x)\MM^{-1}&=U^T\phi(sx)+\pi K^{-1}\chi,
\end{align}
where $U\in GL_N(\mathbb{Z})$ (unimodular), $s=\pm1$ and $\chi\in\mathbb{R}^N$. Hence, each $\MM$ is characterized by the data $(U,~s,~\chi)$.
For example, the data $(\pm I,~1,~0)$ characterize the trivial transform $\mathcal{I}$ and the particle-hole conjugation $\mathcal{C}$ respectively.
%The transformation $\MM^{-1}$ characterized by $(U^{-1},~s,~-sU\chi)$ is the inverse of $\MM$.
While any $\MM$ with $s=1$ acts in an on-site manner, any $\MM$ with $s=-1$ involves a reflection from $x^\mu$ to $\tilde{x}^\mu$ and is called a parity transformation.
The operation $\MM$ implies a transform on the particle operators,
% in Eqs. (\ref{WO}) and (\ref{EO}),
\begin{align}\label{symop}
\MM W^l_\gamma\MM^{-1}&=W^{Ul}_{\tilde{\gamma}},\\
\MM \psi^l(x)\MM^{-1}&=\psi^{Ul}(sx)~e^{i\pi l^TK^{-1}\chi}.
\end{align}
Such transformation naturally induces a map on the quasi-particles, that is, $\MM_{\mathbb{Z}^{N}}:\mathbb{Z}^{N}\rightarrow \mathbb{Z}^{N};l\mapsto Ul.$
%\begin{align}\label{symquasi}
%\MM_{\mathbb{Z}^{N}}:\mathbb{Z}^{N}\rightarrow \mathbb{Z}^{N}~;~~l~\mapsto~Ul.
%\end{align}
Since $U$ is a unimodular matrix, its inverse $U^{-1}$ exists and is also unimodular.
Hence such mapping is an automorphism on the quasi-particle lattice $\mathbb{Z}^{N}$.

%If the state $K$ is equipped with the integer vector $t$, the physical meaning of some transformations becomes clear. For example, the transformation $\mathcal{G}$ characterized by $(I,~1,~\frac{\Phi}{\pi}t)$ corresponds to the $U(1)$ gauge transformation.
%The operation $\mathcal{C}$ specified by $(-I,~1,~0)$ flips the quantum number $q^l_t$ for any choice of $t$; it is the particle-hole transformation or the charge conjugation.

With the understanding of the action of the operations, we are ready to define the symmetry of a state. Any abelian state $K$ is symmetric under $\MM$ if the bulk and edge actions are invariant
under the transformation (\ref{sym}).
Such condition translates to the equation
\begin{align}\label{symK}
sUKU^T=K,
\end{align}
which is independent on the vector $\chi$. If the state $K$ is symmetric under $\MM$, then $\MM$ is called a symmetry of $K$.
%Such symmetry can be broken at the edge in quantum level for non-trivial states \cite{SPTANOMALY3, SPTANOMALY4, SPTANOMALY1, SPTANOMALY2}.
Note that any state $K$ possesses on-site symmetries $\mathcal{I}$ and $\mathcal{C}$. However, the existence of parity symmetry is not guaranteed. If parity symmetry $\MM$ exists, the signature for $K$ must vanish
\footnote{If parity symmetry exists, there is a unimodular $U$ such that $-UKU^T=K$. It implies that $\textrm{sgn}K=-\textrm{sgn}K$ and hence the signature of $K$ vanishes.},
meaning that there is an equal number of left movers and right movers at the edge. Generally, parity symmetry, if exists, may not be unique and any two parity symmetries differ by an on-site symmetry. Besides, if parity symmetry $\MM$ exists, it must be of even order
\footnote{If there is a parity symmetry with odd order $k$, then there is a unimodular $U$ such that $-UKU^T=K$ with $U^k=I$. Then $K=U^kKU^{Tk}=-K$ which implies that K vanishes.},
or otherwise the $K$ matrix vanishes identically. In the paper, unless otherwise specified, we consider
parity symmetries of order two.

%In later sections, unless otherwise specified, we consider parity symmetries of order two.

If $\MM$ is a symmetry of a state $K$, then the automorphism $\MM_{\mathbb{Z}^{N}}$ on the lattice $\mathbb{Z}^{N}$ descends to a well-defined automorphism $\MM_{\AA}$ on the anyon lattice $\AA$,
\begin{align}\label{symanyon}
%\MM_{\AA}:\AA\rightarrow\AA~;~~l~~\textrm{mod}~K\Lambda~\mapsto~Ul~~\textrm{mod}~K\Lambda.
\MM_{\AA}:\AA\rightarrow\AA~;~~l~\mapsto~Ul.
\end{align}
Since $\MM_{\AA}$ is an automorphism on the anyon lattice $\AA$, it can be thought of as a relabeling of the anyons. Under such relabeling $\MM_{\AA}$, the $S$ and $T$ matrix transform as
\begin{align}\label{STtrans}
S_{UlUl'}&=
\left\{\begin{array}{cll}
S_{ll'}&\textrm{if}~~~s=1\\
S_{ll'}^{*}&\textrm{if}~~~s=-1\\
\end{array}
\right.,
\nonumber \\
%~~~~\mbox{and}~~~~
T_{UlUl'}&=
\left\{\begin{array}{cll}
T_{ll'}&\textrm{if}~~~s=1\\
T_{ll'}^{*}&\textrm{if}~~~s=-1\\
\end{array}
\right..
\end{align}
It is known that $S$ and $T$ matrices measure the counter-clockwise statistics whereas $S^*$ and $T^*$ matrices measure the clockwise statistics. Under a parity symmetry $\MM$, orientation of braiding and exchange is flipped due to the parity operation. Hence a full parity symmetry transformation of $S$ and $T$ matrices is a relabeling followed by a complex conjugation. Therefore, $S$ and $T$ matrices are preserved under any symmetry $\MM$. In other words, any symmetry $\MM$ of the topological state $K$ is a symmetry of the anyon statistics.

%Since any symmetry $\MM$ transforms the anyons, it also transforms the physical quantity $q^l_{t}$. Any integer vector $t_c$ is compatible with a symmetry $\MM$ if $Ut_c=st_c$. Similarly, any integer vector $t_i$ is incompatible with a symmetry $\MM$ if $Ut_i=-st_i$. Under the symmetry $\MM$, we have
%\begin{align}\label{qtrans}
%q^{Ul}_{t_c}=q^l_{t_c}
%~~~~\mbox{and}~~~~
%q^{Ul}_{t_i}=-q^l_{t_i}.
%\end{align}
%Hence any symmetry-compatible physical quantity $q^l_{t_c}$ is preserved,
%whereas any symmetry-incompatible physical quantity $q^l_{t_i}$ is flipped under the symmetry $\MM$ of the state $K$. Besides, if an integer vector $t$ is compatible or incompatible with a parity  symmetry, we have the hall conductance $\sigma^{xy}_t$ vanishes.

\subsection{Symmetric Anyons}\label{S2C}

In this subsection, we introduce the notion of symmetric anyons, which show up later when we twist the phase by the symmetry $\MM$. If $\MM$ is a symmetry of the state $K$, we have a group automorphism $\MM_{\AA}$ on the anyon lattice. In the anyon lattice, we have a subgroup consisting of symmetric anyons which are invariant under $\MM_{\AA}$,
\begin{align}
\SS=\{~a\in\AA~|~\MM_{\AA}(a)=a~\}.
\end{align}
%which is just the subgroup $\textrm{Ker}(I-U)$.
While viewed in the integer lattice $\mathbb{Z}^N$, any symmetric anyon $a$ is only changed by a local particle under the relabeling, that is, $Ua=a~\textrm{mod}~K\Lambda$. Let $u_i\in \mathbb{Z}^N$ and $u=\{u_1,~u_2,\dots,~u_M\}$ be a set of generators for $\SS$ such that any $a\in\SS$ can be written as $a=a_iu_i$ mod $K\Lambda$ where $a_i\in\mathbb{Z}$. Then we can rewrite the subgroup as
\begin{align}\label{symsubgroup}
\SS
%=(u\mathbb{Z}^M+K\mathbb{Z}^N)/K\mathbb{Z}^N
=u\mathbb{Z}^M/K\mathbb{Z}^N,
\end{align}
%It is actually a set of equivalent classes instead of a quotient.
where $(u)_{Ii}=(u_i)_I$ is a $N\times M$ matrix with column vectors given by the generators
\footnote{While the anyon lattice $\AA$ is isomorphic to $K^{-1}\mathbb{Z}^N/\mathbb{Z}^N$, the subgroup $\SS$ of symmetric anyons is isomorphic to $K^{-1}u\mathbb{Z}^M/\mathbb{Z}^N$}.
By a slight abuse of notation, we refer, by a generating set, to a matrix of column vectors.
%Let $v$ be the integer eigenbasis of $U$ with eigenvalue $+1$. The group $V$ generated by $v$ is trivially a subgroup of $\SS$.
%The group $V$ generated by $v$ corresponds to the symmetric anyons which is strictly unchanged under $U$ and hence we have $V\leq\SS\leq\AA$
Albeit all the symmetric anyons can be distinguished statistically in $\AA$, some of them statistically look the same in $\SS$. Take any $a,~a'\in\SS$, we define
\begin{align}
a\sim a'
~~\mbox{if}~~
%\theta_{aa''}=\theta_{a'a''}~~\textrm{mod}~2\pi~~\forall a''\in\SS
\theta_{aa''}=\theta_{a'a''}~~\forall a''\in\SS,
\end{align}
where the equal sign for the statistical angles is defined up to mod $2\pi$. It gives a equivalent relation which relates symmetric anyons having the same statistics in the subgroup $\SS$.  Identifying symmetric anyons having the same braiding statistics in subgroup $\SS$, we have the group of distinguishable symmetric anyons in $\SS$ to be
\begin{align}
\hat{\SS}=\SS/\sim.
\end{align}
To make it clear, we define the group homomorphism $f_{u}|_\SS:\SS\rightarrow\mathbb{Q}^{M}/\mathbb{Z}^{M};a\mapsto uK^{-1}a$. In terms of the group homomorphism, the equivalent relation can be written as $a\sim a'$ if $a-a'\in\textrm{Ker} f_u|_{\SS}$ and hence $\hat{\SS}=\SS/\textrm{Ker} f_u|_{\SS}$ is a quotient group. By the fundamental theorem of homomorphisms, the group $\hat{\SS}$ is isomorphic to $f_u|_{\SS}[\SS]=u^TK^{-1}u\mathbb{Z}^M/\mathbb{Z}^M$. The number of distinguishable symmetric anyons in $\SS$ is simply given by the number of lattice points in such lattice \footnote{While $S$ and $\hat{\SS}$ generally depends on the choice of $\MM$, $|S|$ and $|\hat{\SS}|$ are invariant under conjugation of $\MM$ by an on-site symmetry.}. If $\hat{\SS}$ saturates $\SS$, then all the symmetric anyons are statistically distinct, meaning that the subgroup $\SS$ itself forms a modular theory \cite{ANYONSKIT}.

%It should be noted that for any $a\in\SS$, we can write $Ua=a+K\Lambda$ for some integer vector $\Lambda$. Since $a$ is defined in the quotient group $\AA$, the integer vector $\Lambda$ is also living in a quotient group. Let $a'=a+K\lambda$, then we have $Ua'=a'+K\Lambda'$, where $\Lambda'=\Lambda+(sU^{T-1}-I)\lambda$. Hence we get $\Lambda\in\mathbb{Z}^N/(sU^{T-1}-I)\mathbb{Z}^N$.

\section{Twisting by a Symmetry}\label{S3}

Here we consider cutting and gluing a state $K$ along the $x$-axis. Along the cut, we have gapless modes contributed from the two edges. It is known that by adding a gapping term, we can impose a boundary condition on the quasi-particle operators at the interface. By suitably choosing the gapping term that couples the two edges, we can create a defect branch cut which glues the cut. In particular, the gapping term construction for parity branch cut gives us a microscopic view on the construction topological states defined on non-orientable surfaces.

\subsection{The Defect Branch Cut}

\begin{figure}[t]
\vspace{2pc}
\centering
\includegraphics[scale=0.3]{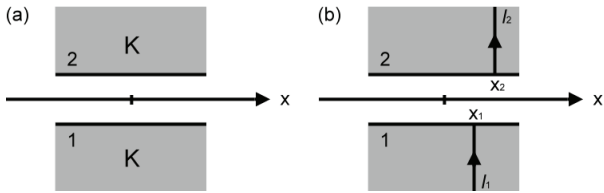}
\caption{(a) The state $K$ is cut along the $x$-axis. The edge $1$ and edge $2$ as a whole is considered as a combined edge. Physically, the two constituent edges do not talk to each other. (b) Without adding any gapping potential coupling the two edges along the combined edge, a Wilson line in the bulk can end at any position on the two edges.}\label{fig1}
\end{figure}

Imagine cutting a state $K$ along the $x$-axis (Fig.\ \ref{fig1}(a)).
The cut can be physically considered as a combined edge along the $x$-axis. We are going to show that, associated to each symmetry $\MM$ of the state $K$, we have a gapping term $\Delta\LL_\MM$ for the combined edge that couples the two constituent edges by introducing hopping of local particles with a twist by the symmetry $\MM$. Denote the lower and the upper region as $1$ and $2$ respectively.
For the edges $1$ and $2$, we have variables $\phi_1$ and $\phi_2$ respectively, which are described by the lagrangians
\begin{align}
\label{edgeslagrangian}
\LL_1&=-\frac{1}{4\pi}\PP_x\phi_1^T(x)K\PP_t\phi_1(x)\\
\LL_2&=s\frac{1}{4\pi}\PP_x\phi_2^T(sx)K\PP_t\phi_2(sx).
\end{align}
Without coupling the fields on the two edges, Wilson lines on one side do not talk to those on the other side and hence can end at any point on the edges (Fig.\ \ref{fig1}(b)).
Let $\Phi(x)=(\phi_1(x),~\phi_2(sx))^T$, the combined edge along the $x$-axis is described by the total lagrangian
\begin{align}
\LL=\LL_1+\LL_2=\frac{1}{4\pi}\PP_x\Phi^T(x)\mathcal{K}\PP_t\Phi(x),
\end{align}
where $\mathcal{K}=-K\oplus sK$ is the enlarged K-matrix with dimension $2N$.
Let $e_I$ be an $N$ components vector with the only non-zero component at the $I$-th entry and equals one. Take $\Lambda_I=(e_I,~U^{T-1}e_I)^T$, since $sUKU^T=K$,
\begin{align}
\Lambda_I^{T}\mathcal{K}\Lambda_J=-e_I^TKe_J+se_I^TU^{-1}KU^{T-1}e_J=0,
\end{align}
for any $I,~J=1,~2,\dots,~N$. Hence $\{\Lambda_I\}$ satisfies the Haldane gapping criterion and is hence a possible choice of null vectors for $\mathcal{K}$ \cite{NULL}.
So corresponding to the symmetry $\MM$, we can construct a gapping potential
\begin{widetext}
\begin{align}\label{gapping}
\Delta\LL_\MM
&=\sum_I -2t\cos\Big[\Lambda_I^T\mathcal{K}\Phi(x)-\pi e_I^T\chi\Big]
=\sum_I -2t\cos\Big[e_I^TK(\phi_1(x)-U^T\phi_2(sx)-\pi K^{-1}\chi)\Big]
\nonumber\\
%&=\sum_I -t\Big[e^{-i\pi e_I^T\chi}\psi_{2}^{UKe_I\dag}(sx)\psi_{1}^{Ke_I}(x)
%+e^{i\pi e_I^T\chi}\psi_{1}^{Ke_I\dag}(x)\psi_{2}^{UKe_I}(sx)\Big],\nonumber\\
&=\sum_I -t\Big[(\MM\psi_2^{Ke_I}(x)\MM^{-1})^\dag\psi_1^{Ke_I}(x)
+\psi_1^{Ke_I}(x)^\dag(\MM\psi_2^{Ke_I}(x)\MM^{-1})\Big],
\end{align}
which is actually a hopping term for the local particles across the constituent edges of the cut. While hopping from the lower edge to the upper edge, $\Delta\LL_\MM$ twists the local particles by the symmetry $\MM$. We will see later that $\Delta\LL_\MM$ also twists anyons by the symmetry $\MM$, and hence the gapping term $\Delta\LL_\MM$ defines the defect branch cut for the symmetry $\MM$ \cite{GappedDW}.

Next, we are going to show that the gapping term $\Delta\LL_\MM$ is symmetric under the symmetry operation $\MM$ if $U$ is orthogonal. By using the cosine expression of the gapping term in Eq.\ ({\ref{gapping}}), we have the transformation
\begin{align}
\MM\Delta\LL_\MM \MM^{-1}
&=\sum_I -2t\cos\Big[e_I^TK(\MM\phi_1(x)\MM^{-1}
-U^T\MM\phi_2(sx)\MM^{-1}-\pi K^{-1}\chi)\Big]
\nonumber\\
&=\sum_I -2t\cos\Big[e_I^TKU^T(\phi_1(sx)-U^T\phi_2(x)-\pi K^{-1}\chi)\Big]
\nonumber\\
&=\sum_I -2t\cos\Big[se_I^TU^{-1}K(\phi_1(x)-U^T\phi_2(sx)-\pi K^{-1}\chi)\Big],
\end{align}
\end{widetext}
where we changed the variable from $x$ to $sx$ in the last step. Since $U$ is orthogonal, that the image of the basis $\{e_1,~e_2,\dots,~e_N\}$ under $U$ is just the basis itself up to some sign changes of the basis vectors. Because sign is not important in the argument of cosine, we have
\begin{align}
\MM\Delta\LL_\MM \MM^{-1}=\Delta\LL_\MM.
\end{align}
So $\Delta\LL_\MM$ is symmetric under the symmetry $\MM$, adding the defect branch cut does not break the symmetry $\MM$ of the whole system. In particular, the matrix $U$ can always be chosen as an orthogonal matrix for any parity symmetry of order two, hence such parity branch cut is always symmetry preserving.

\subsection{Boundary Condition Across the Branch Cut}\label{S3B}

Here, we derive the boundary condition for the Wilson operators across the branch cut imposed by the gapping term $\Delta\LL_\MM$. By coupling the two edges, $\Delta\LL_\MM$ connects the Wilson lines on both sides of the interface. We are going to show that the gapping term $\Delta\LL_\MM$ twists the anyons by the symmetry $\MM$ across the branch cut. Since the gapping term $\Delta\LL_\MM$ pins the argument of the cosine term to the local minimum of minus cosine, quantum mechanically, the possible eigenvalues of the operator,
\begin{align}\label{guass}
G(x):=\phi_1(x)-U^T\phi_2(sx)-\pi K^{-1}\chi,
\end{align}
is given by $2\pi K^{-1}p$, where $p$ is an anyon carried by a Wilson line along the defect branch cut \cite{ANYSYM}. Pick an eigenstate with fixed $p$, we have the following equation
\begin{align}
\psi_2^{Ul\dag}(sx)\psi_1^{l}(x)=e^{(i\pi l^TK^{-1}\chi+i2\pi l^TK^{-1}p)}.
\end{align}
Since $\langle\psi_2^{Ul\dag}(sx)\psi_1^{l}(x)\rangle\neq0$, it describes a particle condensate of $-l$ at $x$ on edge $1$ and $Ul$ at $sx$ on edge $2$ along the interface \cite{DEFECT,PROEDGE}. Physically, it should be expected that an anyon $l$ getting to edge $1$ from below at $x$ can fuse with a condensed particle pair to becomes an anyon $Ul$ getting away from edge $2$ at $sx$.
%Physically, while getting across the branch cut from below, in addition to the phase $e^{i2\pi l^TK^{-1}p}$ contributed by the Wilson line along the interface, the anyon $l$ also acquire a phase $e^{i\pi l^TK^{-1}\chi}$ from the branch cut defined by $\Delta\LL_\MM$.

\begin{figure}[t]
\centering
\includegraphics[scale=0.3]{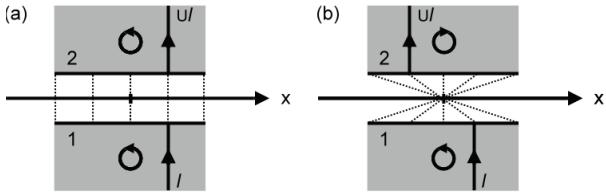}
\caption{ The gapping term $\Delta\LL_\MM$ glues the two edges by imposing boundary condition on the Wilson operators crossing the branch cut. Essentially, it twists anyons by the symmetry $\MM$ across the branch cut. (a) The gapping term for an on-site symmetry glues the edges directly with an anyonic twist $U$. (b) The gapping term for a parity symmetry glues the edges with a spatial twist together with an anyonic twist $U$.}\label{fig2}
\end{figure}

Consider Wilson operators which do not involve any Wilson line along branch cut. In such case, $p$ is fixed and any physical operator must not change the eigenvalue of $G(x)$. More precisely, any physical operator must commute with $G(x)$. It can be easily checked that, without a partner on the other side, any Wilson line on one side ending on the edge does not commute with $G(x)$. Consider two Wilson lines, one on each side, ending on the edges $1$ and $2$ respectively as indicated in Fig.1. Pulling back to the edges, the Wilson operator becomes the field operator $\psi_2^{l_2\dag}(x_2)\psi_1^{l_1}(x_1)$.
For such Wilson operator to be physical, we have the condition that
\begin{align}
&
[\psi_2^{l_2\dag}(x_2)\psi_1^{l_1}(x_1),G(x)]=0
%&~~~\Leftrightarrow~~~
%[i(-l_2^T\phi_2(x_2)+l_1^T\phi_1(x_1)),G(x)]=0\nonumber\\
%&~~~\Leftrightarrow~~~
%i[l_2^T\phi_2(x_2),U^T\phi_2(sx)]+i[l_1^T\phi_1(x_1),\phi_1(x)]=0\nonumber\\
%&~~~\Leftrightarrow~~~
%\pi l_2^TK^{-1}U\textrm{sgn}(x_2-sx)-\pi l_1^TK^{-1}\textrm{sgn}(x_1-x)=0\nonumber\\
%&~~~\Leftrightarrow~~~
%\pi sl_2^TU^{T-1}\textrm{sgn}(x_2-sx)-\pi l_1^T\textrm{sgn}(x_1-x)=0\nonumber\\
%&~~~\Leftrightarrow~~~
%\pi l_2^TU^{T-1}\textrm{sgn}(sx_2-x)-\pi l_1^T\textrm{sgn}(x_1-x)=0\nonumber\\
%&~~~\Leftrightarrow~~~
%-\pi l_2^TU^{T-1}\textrm{sgn}(x-sx_2)+\pi l_1^T\textrm{sgn}(x-x_1)=0\nonumber\\
\nonumber \\
~~~\Leftrightarrow~~~&
l_{2}\textrm{sgn}(x-sx_2)-Ul_{1}\textrm{sgn}(x-x_1)=0\nonumber\\
~~\Leftrightarrow~~~&
l_{2}=Ul_{1}~~\mbox{and}~~x_2=s x_1.
\end{align}
Pictorially, while crossing the interface from below, any anyon $l$ is twisted by $U$, at the same time, the anyon trajectory goes from $x$ to $sx$
\footnote{One can also check that while going cross the interface from above, $\Delta\LL_\MM$ do the inverse, meaning that $l$ is twisted by $U^{-1}$ and the anyon trajectory goes from $x$ to $sx$.}.
For any on-site symmetry $\MM$, the anyon trajectory is unaffected across the branch cut, so the gapping term $\Delta\LL_\MM$ glues the upper side and the lower side directly with a twist $U$ of anyon label (Fig.\ \ref{fig2}(a)). So loop orientation is preserved across any on-site defect branch cut. The gapping term $\Delta\LL_\mathcal{I}$ introduces a trivial gluing and corresponds to a direct hopping term across the interface. The gapping term $\Delta\LL_\mathcal{C}$ condenses particle pairs which is equivalent to adding a SC wire on the interface. For any parity symmetry $\MM$, the anyon trajectory is twisted from $x$ to $-x$ across the branch cut, hence the gapping term $\Delta\LL_\MM$ glues the upper side and the lower side with a spatial twist in addition to the anyonic twist $U$ (Fig.\ \ref{fig2}(b)).
%So loop orientation get flipped across any parity branch cut. $\Delta\LL_\MM$ for any parity symmetry corresponds to a non-local hopping term across interface, tunneling particles with a relabeling $U$.
Generally, the change in loop orientaion together with the anyon relabelling ensure that the anyon statistics is locally well-defined across any branch cut.

The discussion above concerns the boundary condition for the Wilson operators crossing the branch cut. Here we remark on the boundary condition for $a_\mu$ implied by the gapping term $\Delta\LL_\MM$. From Eq.\ (\ref{guass}), we have
\begin{align}
\phi_1(x)&=\MM\phi_2(x)\MM^{-1}+2\pi K^{-1}p,
\end{align}
which relates the edge variables $\phi_1$ and $\phi_2$ at the interface.
By differentiating such equation w.r.t. $x$, we get the boundary condition for $a_x$ across the branch cut, that is
\begin{align}\label{twistedax}
a_{1x}(x)&=\MM a_{2x}(x)\MM^{-1}.
\end{align}
Hence the branch cut defined by $\Delta\LL_\MM$ twists $a_x$ by $\MM$ from edge $2$ to edge $1$. In other words, the gapping term $\Delta\LL_\MM$ imposes a twisted boundary condition for the bulk gauge field $a_\mu$ at the defect branch cut.

\section{States on the Non-orientable Surfaces}\label{S5}

In this section, we are going to construct topological states on non-orientable surfaces. We construct a twisted ribbon by adding a defect branch cut on a ribbon geometry along the axial direction and show that the twisted ribbon is a M\"obius strip if the defect branch cut corresponds to a parity symmetry. On the M\"obius strip, we see that only the parity symmetric anyons are left. To proceed, we define topological states on the real project plane by compactifying the M\"obius strip. Using the real project plane as the generator, we can define topological field theories on any non-orientable closed surface. We construct the ground states on arbitrary non-orientable closed surface and obtain the ground state degeneracy (GSD). Using the Dyck's theorem, we then discuss the robustness of such GSD by showing that it is invariant under any smooth deformation of the surface.

\subsection{ The Twisted Ribbon}

In this section, we construct the twisted ribbon which is a ribbon with a defect branch cut along the axial direction and it gives a M\"obius strip if the branch cut corresponds to a parity symmetry. We discuss the physical Wilson loop operators generally on the twisted ribbon, showing that only the symmetric anyons are relevant under the twist. Then we discuss the physical properties of the parity symmetric anyons on the M\"obius strip.

% and show that their statistical properties are dictated by the non-orientability of the M\"obius strip.
%Note that the twisted ribbon has an edge/edges, and supports gapless excitations there. We do not discuss the properties of edge states in this paper.

\subsubsection{Physical Wilson Loops Operators}

\begin{figure}[t]
\centering
\includegraphics[scale=0.3]{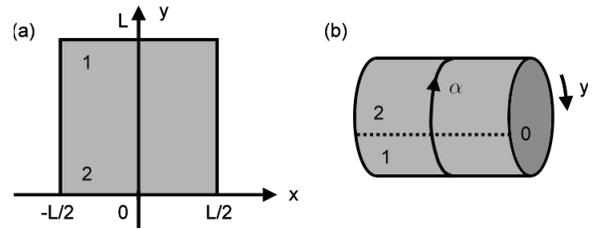}
\caption{
(a) The state $K$ is prepared in a square shape with length scale $L$. The upper and lower edges are labeled by $1$ and $2$ respectively. (b) The twisted ribbon is constructed by passing the upper edge through the back and gluing it to the lower edge using the gapping term $\Delta\LL_\MM$. If $\Delta\LL_\MM$ is on-site, the twisted ribbon is a cylinder $\mathbb{C}$; If $\Delta\LL_\MM$ corresponds to a parity symmetry, it is a M\"obius strip $\mathbb{M}$}\label{fig3}
\end{figure}

Consider a state $K$ with a symmetry $\MM$. The state is prepared in a square shape with width $L$ in $x$-direction and length $L$ in $y$-direction (Fig.\ \ref{fig3}(a)). Edge $1$ at $y=L$ is passed through the back of the state and glued to the edge $2$ at $y=0$ by using the gapping term $\Delta\LL_{\MM}$. We call the resulting ribbon a ribbon twisted by the symmetry $\MM$. Consider dragging a loop along the non-trivial cycle of the twisted ribbon. If $\Delta\LL_\MM$ is on-site, the orientation of the loop is unchanged after a round trip, hence the twisted ribbon is topologically the same as a cylinder $\mathbb{C}$
\footnote{It should be noted that $\mathbb{C}$ does not stand for the set of complex numbers in this paper.}. On the other hand, if $\Delta\LL_\MM$ is a gapping term for parity, then the loop orientation is flipped under a round trip and therefore the twisted ribbon is a M\"obius strip $\mathbb{M}$. The two surfaces share the same first homology group,
\begin{align}
H_1(\mathbb{C})=H_1(\mathbb{M})=\langle~\alpha~\rangle=\mathbb{Z},
\end{align}
where $\alpha$ is a single cycle going in positive $y$-direction. A loop is either {\it one-sided} or {\it two-sided}, depending on whether its regular neighborhood is an annulus or a M\"obius strip. The cycle $\alpha$ in the cylinder $\mathbb{C}$ is two-sided whereas the cycle $\alpha$ in the M\"obius strip $\mathbb{M}$ is one-sided.

%Here, we discuss the set of physical Wilson loop operators on the twisted ribbons $\mathbb{C}$ and $\mathbb{M}$.

On the twisted ribbon, any physical Wilson loop operator must have its final anyon the same as its initial anyon in the anyon lattice. We are going to show the that set of all physical Wilson loop operators is precisely
\begin{align}\label{allWil}
\mathcal{W}=\{~W^a_\alpha~|~a\in\SS~\}.
\end{align}
In other word, only the symmetric anyons are relevant. Under the operation $W^a_\alpha\ast W^{a'}_\alpha=W^{a+a'}_\alpha$, which is the multiplication of Wilson operators up to a trivial phase,
$\mathcal{W}$ forms a group isomorphic to the subgroup $\SS$.

Notice that all operators in $\mathcal{W}$ are physical, so it suffices to show that any physical Wilson loop operator can be written as an element in $\mathcal{W}$. Any Wilson operator going along a trivial cycle is physical and can be represented by the trivial element $W^0_\alpha$ in $\mathcal{W}$. Any Wilson operator going through the cycle $\alpha$ by $k$ times can be written as $W^l_\alpha\ast W^{Ul}_\alpha\ast\dots\ast W^{U^{k-1}l}_\alpha
=W^{(I+U+\dots+U^{k-1})l}_\alpha$.
%\begin{align}
%W^l_\alpha\ast W^{Ul}_\alpha\ast\dots\ast W^{U^{k-1}l}_\alpha
%=W^{(I+U+\dots+U^{k-1})l}_\alpha.
%\end{align}
In order for such Wilson operator to form a loop physically,
the initial and the final anyons must be the same in the anyon lattice.
Hence we have the requirement that $U^kl=l~\textrm{mod}~K\Lambda$ which implies $(I+U+\dots+U^{k-1})l\in\SS$.
%\begin{align}
%&
%\quad
%U(I+\dots+U^{k-2}+U^{k-1})l
%\nonumber \\
%&\quad =(U+\dots+U^{k-1}+I)l~\mod K\Lambda
%\nonumber\\
%&\Rightarrow~(I+U+\dots+U^{k-1})l\in\SS.
%\end{align}
Therefore, any  physical Wilson loop operator going along the cycle $\alpha$ by $k$ times is also an element in $\mathcal{W}$. To conclude, the set of all possible physical Wilson loop operators is precisely given by the abelian group $\mathcal{W}$.

\subsubsection{The Parity Symmetric Anyons}\label{ParitySym}

On the M\"obius strip $\mathbb{M}$, the only relevant anyons are the anyons invariant under the parity symmetry, which we call the parity symmetric anyons. Mathematically, we can always find a M\"obius strip $\mathbb{M}$ on any arbitrary non-orientable surface. Hence parity symmetric anyons show up on any non-orientable surface. Consequently, the topological properties of a topological state defined on non-orientable surfaces are deeply related to these parity symmetric anyons. Here we discuss the properties of these parity symmetric anyons on $\mathbb{M}$ and the structure of the subgroup $\SS$.

Firstly, the statistics of the parity symmetric anyons exhibits a loss of orientation. Under the parity symmetry, recall that the $S$ and $T$ matrices are invariant under the anyon relabeling followed by a complex conjugation which takes into account the change of orientation of statistics. Notice that the $S$ and $T$ matrices while they are restricted to $\SS$ are invariant under the relabeling, so
\begin{align}\label{statid}
S_{aa'}=S_{aa'}^*
~~~~\mbox{and}~~~~
T_{aa'}=T_{aa'}^*,
\end{align}
where the equation for the T-matrix is defined up to a sign for systems which involve local fermions. Physically, it means that the parity symmetric anyons have their left hand and right hand statistics identified, exhibiting a loss of orientation. By the definition in Eq.\ (\ref{ST}), we get a constraint on their mutual-statistics
\begin{align}\label{statpro}
\theta_{aa'}=0~\textrm{or}~\pi~~\textrm{mod}~2\pi.
\end{align}
Hence the parity  symmetric anyons are either mutually bosonic or semionic. Similarly, using Eq.\ (\ref{ST}), we get a constraint on their self-statistics
\begin{align}
~~~~\theta_{a}=
\left\{\begin{array}{cl}
0~\textrm{or}~\pi~~\textrm{mod}~2\pi,&\textrm{for bosonic system;}\\
0~\textrm{or}~\frac{\pi}{2}~~\textrm{mod}~~\pi,&\textrm{otherwise.}\\
\end{array}
\right.
\end{align}
For a bosonic system, any parity symmetric anyon is either a self-boson or a self-fermion. If the microscopic hamiltonian involves local fermions, we may have self-semionic parity symmetric anyons. More pictorially, imagine braiding or spinning a parity symmetric anyon at difference points along the one-sided cycle $\alpha$ in the M\"obius strip $\mathbb{M}$. While it is carried back to its original position, the anyon label is unchanged whereas the orientation of the braiding or spinning is flipped. Since locally we have a well-defined statistics, the initial statistics must agree with the final statistics. So the counter-clockwise statistics merges with the clockwise statistics for the parity symmetric anyons.

%Besides, $\theta_{aa'}=0~\textrm{mod}~2\pi$ and $\theta_{a}=0~\textrm{mod}~2\pi/\pi$, $\forall a,~a'\in V$ \footnote{Given a parity symmetry. Take two parity symmetric anyons $a,~a'\in V$, we have $a^TK^{-1}a'=a^TU^TK^{-1}Ua'=-a^TK^{-1}a'$, which implies that $a^TK^{-1}a'$ vanishes}.

Secondly, there is an explicit formula for the number of distinguishable parity symmetric anyons in $\hat{\SS}$. Denote $[~]$ sandwiching a matrix as taking mod $2$ to the entries. Let $u$ be a generating set for $\SS$
\footnote{The result is independent on the choice of the generating set $u$ for $\SS$. We can as well take the generating set $u$ as the whole group $\SS$ where the matrix in the exponent shown in (Eq.\ \ref{dissymsubgroup}) times $\pi$ is simply the mutual-statistical angle $\theta_{aa'}$ between the parity symmetric anyons in $\SS$}, the order of $\hat{\SS}$ is
\begin{align}\label{dissymsubgroup}
|\hat{\SS}|=2^{\textrm{Rk}[2u^TK^{-1}u]}.
\end{align}
Roughly speaking, each independent column vector in $[2u^TK^{-1}u]$ spans two states. To show this, notice that $\hat{\SS}$ is isomorphic
\footnote{$\hat{\SS}\simeq f_u|_{\SS}[\SS]=u^TK^{-1}u\mathbb{Z}^M/\mathbb{Z}^M\simeq2u^TK^{-1}u\mathbb{Z}^M/2\mathbb{Z}^M=[2u^TK^{-1}u]\mathbb{Z}^M_2$}
to the lattice $[2u^TK^{-1}u]\mathbb{Z}^M_2$, it suffices to count the number of lattice point in such lattice. From Eq.\ (\ref{statpro}), the entries of the matrix $[2u^TK^{-1}u]$ are in $\mathbb{Z}_2$. We first write such matrix in its Smith normal form $R$ which is diagonal and with entries also in $\mathbb{Z}_2$. Since each diagonal entry of $R$ with value $1~\textrm{mod}~2$ gives two lattice points, the total number of lattice points is $2^{\textrm{Rk}(R)}$. Since Smith normal form is rank-preserving, Eq.\ (\ref{dissymsubgroup}) follows.

%\begin{align}
%&\SS=u\mathbb{Z}^M/K\mathbb{Z}^N=u'\mathbb{Z}^{M'}/K\mathbb{Z}^N\nonumber\\
%&\Rightarrow K^{-1}u\mathbb{Z}^M/\mathbb{Z}^N=K^{-1}u'\mathbb{Z}^{M'}/\mathbb{Z}^N\nonumber\\
%&\Rightarrow K^{-1}u\mathbb{Z}^M+\mathbb{Z}^N=K^{-1}u'\mathbb{Z}^{M'}+\mathbb{Z}^N\nonumber\\
%&\Rightarrow u^TK^{-1}u\mathbb{Z}^M+u^T\mathbb{Z}^N=u^TK^{-1}u'\mathbb{Z}^{M'}+u^T\mathbb{Z}^N\nonumber\\
%&\Rightarrow u^TK^{-1}u\mathbb{Z}^M/\mathbb{Z}^{M}=u^TK^{-1}u'\mathbb{Z}^{M'}/\mathbb{Z}^{M}
%\end{align}
%\begin{align}
%u'^TK^{-1}u\mathbb{Z}^M/\mathbb{Z}^{M'}=u'^TK^{-1}u'\mathbb{Z}^{M'}/\mathbb{Z}^{M'}
%\end{align}
%\begin{align}
%u^TK^{-1}u\mathbb{Z}^M/\mathbb{Z}^{M}=u^TK^{-1}u'\mathbb{Z}^{M'}/\mathbb{Z}^{M}\simeq u'^TK^{-1}u\mathbb{Z}^M/\mathbb{Z}^{M'}=u'^TK^{-1}u'\mathbb{Z}^{M'}/\mathbb{Z}^{M'}
%\end{align}

%Similarly, consider a topological state $K$ with symmetry $\MM$.
%Suppose the state $K$ is equipped with an integer vector $t_i$ incompatible with $\MM$.
%Take any symmetric anyon $a\in \SS$
%under the symmetry $\MM$,
%we have $q^{Ua}_{t_i}=q^a_{t_i}~\textrm{mod}~1$. Since $q^{Ua}_{t_i}=-q^a_{t_i}$, we get a constraint on the physical quantity $q^a_{t_i}$ carried by $a$,
%\begin{align}
%q^{a}_{t_i}
%=0~\textrm{or}~\frac{1}{2}~~\textrm{mod}~1.
%\end{align}
%So the quantity $q^{a}_{t_i}$ of any symmetric anyon $a\in\SS$ must either be zero or one-half. In addition, we have a more restrictive condition $q^{a}_{t_i}=0~\textrm{mod}~1$, for any $a\in V$.

%The discussion above is general for any given parity symmetry.

Lastly, taking into account the fact that the parity symmetry is of order two, as proven in Appendix\ \ref{S4}, there is an additional structural property for the subgroup $\SS$ of symmetric anyons. If a state $K$ (dim $K=N$) possesses a $\mathbb{Z}_2$ parity symmetry $\MM$, then there exists a set $v$ of $n=N/2$ mutually bosonic integer vectors and a set $w$ of $n'\geq0$ mutually bosonic integer vectors such that
 \begin{align}
u=v\cup w
\end{align}
generates $\SS$. The set $v$ is precisely the integer eigenbasis of the transformation matrix $U$ with eigenvalue $+1$. We call $u=v\cup w$ as the generating set for $\SS$ induced by the $\mathbb{Z}_2$ parity symmetry. In term of such generating set,
\begin{align}\label{statpro2}
u^TK^{-1}u
&=
\left(\begin{array}{cc}
v^TK^{-1}v&v^TK^{-1}w\\
w^TK^{-1}v&w^TK^{-1}w\\
\end{array}
\right)
%\nonumber \\
%&=
%\left(\begin{array}{cc}
%0~\textrm{mod}~1&0~\textrm{or}~1/2~\textrm{mod}~1\\
%0~\textrm{or}~1/2~\textrm{mod}~1&0~\textrm{mod}~1\\
%\end{array}
%\right),
\end{align}
%\begin{align}\label{statpro2}
%\left[2u^TK^{-1}u\right]
%&=
%\left(\begin{array}{cc}
%~[2v^TK^{-1}v]&[2v^TK^{-1}w]\\
%~[2w^TK^{-1}v]&[2w^TK^{-1}w]\\
%\end{array}
%\right)
%\end{align}
where $v^TK^{-1}v$ is a $n\times n$ integer matrix and $w^TK^{-1}w$ is a $n'\times n'$ integer matrix whereas $v^TK^{-1}w=(w^TK^{-1}v)^T$ is a $n\times n'$ matrix with entries living in $\frac{1}{2}\mathbb{Z}$. An immediate observation from Eq.\ (\ref{statpro2}) is that, since sandwiching the matrix $u^TK^{-1}u$ by any integer vector must give an integer, the parity symmetric anyons cannot be self-semionic, meaning that they must be either self-bosonic or self-fermionic. Another observation from Eq.\ (\ref{statpro2}) is that the diagonal blocks of the matrix $[2u^TK^{-1}u]$ vanish and the entries of the off-diagonal blocks are in $\mathbb{Z}_2$. Notice that the rank of $[2u^TK^{-1}u]$ is twice the rank of $[2w^TK^{-1}v]$. From Eq.\ (\ref{dissymsubgroup}), the number $|\hat{\SS}|$ of distinguishable parity symmetric anyons, like the total number of anyons in a $\mathbb{Z}_2$ parity symmetric state, must a square of some integer. Let $V$ and $W$ be the subgroup generated by $v$ and $w$ respectively. Actually, the number of  distinguishable parity symmetric anyons equals the square of the number of anyons in $V$ distinguishable in $W$
\footnote{The anyons in $V$ distinguishable in $W$ can be defined through the equivalent relation: Take $a_v, a_v'\in V$, $a_v\sim a_v'$ if $\theta_{a_va_w}=\theta_{a_v'a_w}~\forall a_w\in W$. It turns out that the number of anyons in $V$ distinguishable in $W$ is $2^{\textrm{Rk}[2w^TK^{-1}v]}$ which equals the number of anyons in $W$ distinguishable in $V$.}. As a final remark on Eq.\ (\ref{statpro2}), the column vectors in $v$ (or $w$) are mutually bosonic, so the anyons in the group $V$ (or $W$) must also be mutually bosonic. This statistical property is very important later when we construct the ground states on non-orientable surfaces.

\subsection{The Real projective plane}\label{RP}

\begin{figure}[t]
\centering
\includegraphics[scale=0.3]{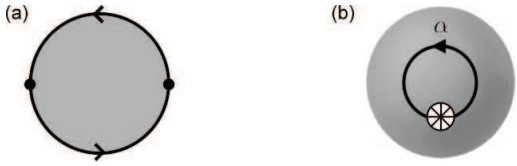}
\caption{(a) The real projective plane $\mathbb{P}$ is constructed by shrinking the edge of the M\"obius strip $\mathbb{M}$ to a point. (b) By continuous deformation, the real projective plane $\mathbb{P}$ can be turned into a cross cap which is a punctured $\mathbb{S}^2$ with anti-nodal points of the hole identified. The cross cap can be viewed as a pair of parity defects on $\mathbb{S}^2$.}\label{fig4}
\end{figure}

Here, we are going to construct a topological state defined on the real projective plane $\mathbb{P}$. Generally, to define topological states on non-orientable surfaces, we glue the surfaces suitably by using the parity branch cuts. Since geometrically gluing with a double reflection is a direct gluing, to inherit this property for the parity branch cut, we require that twisting twice by parity symmetry $\MM$ is trivial, namely $\MM^2=I$.
Note that the M\"obius strip has an edge with length $2L$.
Compactifying the M\"obius strip $\mathbb{M}$ by shrinking the edge to a point (one-point compactification),
we get the real projective plane $\mathbb{P}$, where there is no boundary (see Fig.\ \ref{fig4}(a)).
Alternatively, one can construct the real projective plane $\mathbb{P}$ by starting with the square-shaped topological state shown in Fig.\ \ref{fig3}. The method is to glue the top edge to the bottom edge in antiparallel sense by using the gapping term $\Delta\LL_\MM$ for parity and to glue the right edge to the left edge in antiparallel sense similarly. The two construction of real projective plane are topologically equivalent. By continuously deforming the manifold, we can turn the real projective plane $\mathbb{P}$ to a cross cap
which is a punctured $\mathbb{S}^2$ with anti-nodal points of the hole get identified (see Fig.\ \ref{fig4}(b)). Such cross cap can be viewed as a pair of parity defects connected by a parity branch cut on $\mathbb{S}^2$. Let $\alpha$ be the one-sided loop going through the cross cap once. The first homology group of $\mathbb{P}$ is
\begin{align}
H_1(\mathbb{P})=\langle~\alpha~|~2\alpha=0~\rangle=\mathbb{Z}_2,
\end{align}
where $\alpha$ is the generator and $2\alpha=0$ is the group relation. A loop on $\mathbb{P}$
is two-sided if it gets through the cross cap even number of times.
Since $2\alpha=0$, there is no non-trivial two-sided loop on $\mathbb{P}$. The intersection number of the loop $\alpha$ with itself is $I(\alpha,~\alpha)=1~\textrm{mod}~2$ which is $\mathbb{Z}_2$ valued since the binary operation $I$ is alternating. Due to the intersection of the loop $\alpha$ with itself, we can have a non-trivial Wilson operator algebra on the real projective plane $\mathbb{P}$.

%Here, we are going construct the ground states and calculate the GSD on $\mathbb{P}$.

\begin{figure}[t]
\centering
\includegraphics[scale=0.3]{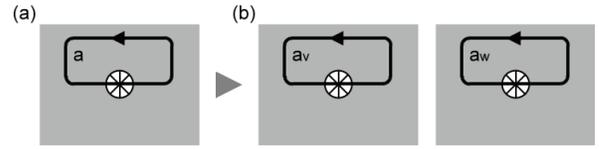}
\caption{(a) The physical Wilson loop operators $W^a_\alpha$ in the group $\mathcal{W}(\mathbb{P})$. (b) By the structural property of the subgroup $\SS$, we can construct generators $W^{a_v}_\alpha$ and $W^{a_w}$ which can be treated as the raising and measuring operators respectively.}\label{fig5}
\end{figure}

To construct the ground states on the real projective plane $\mathbb{P}$, we need to figure out the group of physical Wilson operators $\mathcal{W}(\mathbb{P})$ and then construct a generating set for $\mathcal{W}(\mathbb{P})$ in which the measuring and raising operators can be identified. Since any physical Wilson loop operator going along $\alpha$ must carry a parity symmetric anyon, the set of physical Wilson loop operators is given by
\begin{align}\label{WP}
\mathcal{W}(\mathbb{P})=\{~W^a_\alpha~|~a\in\SS~\}.
\end{align}
forms an abelian group under the fusion operation \footnote{$\mathcal{W}(\mathbb{P})$ forms an abelian group under the fusion operation $W^a_\alpha\ast W^{a'}_\alpha=W^{a+a'}_\alpha$}.
For any abelian state $K$ with parity symmetry $\MM$ of order two,
recall from section \ref{ParitySym} that the set $u=v\cup w$ generates the subgroup $\SS$ of parity symmetric anyons. In other words, for any $a\in\SS$, we can write $a=a_v+a_w$ for some $a_v\in V$ and $a_w\in W$. So the group $\mathcal{W}(\mathbb{P})$ is generated by the Wilson operators $W^{a_v}_\alpha$ and $W^{a_w}_\alpha$,
\begin{align}
\mathcal{W}(\mathbb{P})=\langle~W^{a_v}_\alpha,~W^{a_w}_\alpha~|~a_v\in V,~a_w\in W~\rangle.
\end{align}
We have shown that anyons in $V$ are mutually bosonic and anyons in $W$ are also mutually bosonic. So we can simultaneously diagonalize the operators $W^{a_w}_\alpha$ and treat $W^{a_v}_\alpha$ as the raising operators (see Fig.\ \ref{fig5}). We can as well construct the ground states another way round but the results for the GSD are the same.

Take the measuring operators $W^{a_w}_\alpha$ and the raising operators $W^{a_v}_\alpha$,
we are ready to construct the ground states which corresponds to different fluxes measured by $W^{a_w}_\alpha$.
Let $w_i$ be the $i$-th vector in the generating set $w$. From the Wilson operator algebra, we have the equation
\begin{align}\label{projal}
W^{w_i}_{\alpha}W^{a_v}_{\alpha}=e^{i2\pi w_i^TK^{-1}a_v} W^{a_v}_{\alpha}W^{w_i}_{\alpha},
\end{align}
where $w^TK^{-1}a_v$ is the flux vector measured w.r.t. $w$.
%\textcolor{red}{with number of components given by $n'$.}
The measurement of flux vectors w.r.t $w$ defines a group homomorphism
$f_{w}|_V:V\rightarrow \mathbb{Q}^{n'}/\mathbb{Z}^{n'};a_v\mapsto w^TK^{-1}a_v$.
%\begin{align}
%f_{w}|_V:&
%\quad
%V\rightarrow \mathbb{Q}^{n'}/\mathbb{Z}^{n'};~a_v~\mapsto~w^TK^{-1}a_v.
%%a_v~~\textrm{mod}~K\Lambda~\mapsto~w^TK^{-1}a_v~~\textrm{mod}~\mathbb{Z}^{n'},
%\end{align}
%which is the restriction of the map $f_w$ to the subgroup $V$.
Take a reference state $|0\rangle$ with $W^{w_i}_{\alpha}|0\rangle=|0\rangle$. A quantum state $|a_v\rangle$ is constructed by operating the raising operator $W^{a_v}_{\alpha}$ on the reference state $|0\rangle$. Two states $|a_v\rangle$ and $|a'_v\rangle$ are identified if they give the same flux. So we have
\begin{align}\label{GSP}
&|a_v\rangle=W^{a_v}_{\alpha}|0\rangle
~~\mbox{and}~~\\
&W^{w_i}_{\alpha}|a_v\rangle=e^{i2\pi w_i^TK^{-1}a_v}|a_v\rangle,
\end{align}
where $a_v\in V/\textrm{Ker} f_w|_{V}\simeq f_w|_{V}[V]$. The first equation defines the ground states on the real projective plane $\mathbb{P}$ while the second equation tells the flux measured by $W^{w_i}$ for each state. The quantity GSD$(\mathbb{P})$ can be obtained by simply counting the number of possible flux vectors in the image $f_w|_{V}[V]=w^TK^{-1}v\mathbb{Z}^{n}/\mathbb{Z}^{n'}$.
By using the result in Eq.\ (\ref{statpro2}), it can be shown that
\footnote{Note that $f_w|_{V}[V]=w^TK^{-1}v\mathbb{Z}^{n}/\mathbb{Z}^{n'}\simeq v^TK^{-1}w\mathbb{Z}^{n'}/\mathbb{Z}^{n}$. So $f_w|_{V}[V]^2\simeq w^TK^{-1}v\mathbb{Z}^{n}/\mathbb{Z}^{n'}\times v^TK^{-1}w\mathbb{Z}^{n'}/\mathbb{Z}^{n}\simeq u^TK^{-1}u\mathbb{Z}^{M}/\mathbb{Z}^{M}\simeq \hat{\SS}$, where the second isomorphic relation follows from Eq.\ (\ref{statpro2}).}
$f_w|_{V}[V]^2\simeq\hat{\SS}$, therefore
\begin{align}\label{GSDP}
\textrm{GSD}(\mathbb{P})=|\hat{\SS}|^{1/2},
\end{align}
which is the root of the number of distinguishable parity symmetric anyons. Different ground states simply corresponds to different fluxes measured by $W^{a_w}_\alpha$ under the flux insertion by $W^{a_v}_{\alpha}$. Hence $\textrm{GSD}(\mathbb{P})$ counts the number of anyons in $V$ distinguishable in $W$.

\subsection{The Klein Bottle}

\begin{figure}[t]
\centering
\includegraphics[scale=0.3]{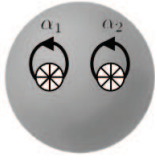}
\caption{The Klein bottle $\mathbb{K}$ can be constructed by a connected sum of two cross caps. $\alpha_1$ and $\alpha_2$ are the one-sided loops associated with the first and the second cross caps respectively. The Klein bottle $\mathbb{K}$ can essentially be viewed as two pairs of parity defects on $\mathbb{S}^2$.}\label{fig6}
\end{figure}

In the previous section, we have constructed a state defined on the real projective plane $\mathbb{P}$. By considering the connected sum of two cross caps, we get a state defined on the Klein bottle $\mathbb{K}$ (Fig.\ \ref{fig6}). It can be viewed as two pairs of parity defects on $\mathbb{S}^2$.
Alternatively, one can construct the Klein bottle $\mathbb{K}$ by suitably gluing the square-shaped topological state shown in Fig.\ \ref{fig3}. The method is to glue the top edge to the bottom edge in antiparallel sense by using the gapping term $\Delta\LL_\MM$ for parity and to glue the right edge to the left edge in parallel sense by using the trivial twist $\Delta\LL_\mathcal{I}$. It can be shown that the arrangements of parity branch cuts from the two constructions are the same.
% iff $\MM^2=I$.
For the sake later of generalization, we stick to the picture where the Klein bottle $\mathbb{K}$ is a connected sum of two cross caps. For each of the two cross caps on the Klein bottle $\mathbb{K}$, we have an associated one-sided loop.
Let $\alpha_1$ and $\alpha_2$ be the one-sided loops going through the first and the second cross caps respectively. The first homology group of the Klein bottle $\mathbb{K}$ is
\begin{align}
H_1(\mathbb{K})=\langle~\alpha_1,~\alpha_2~|~2(\alpha_1+\alpha_2)=0~\rangle=\mathbb{Z}\times\mathbb{Z}_2.
\end{align}
On the Klein bottle $\mathbb{K}$, there are non-trivial two-sided loops. The subgroup of two-sided loops in $H_1(\mathbb{K})$ is generated by the loops $2\alpha_1$, $2\alpha_2$ and $\beta_1=\alpha_1+\alpha_2$. The intersection number between the loops in $H_1(\mathbb{K})$ can be obtained by making use of the bilinear intersection $I(\alpha_i,~\alpha_j)=\delta_{ij}~\textrm{mod}~2$.

%Here, we construct the ground states and calculate the GSD on $\mathbb{K}$.

\begin{figure}[t]
\centering
\includegraphics[scale=0.27]{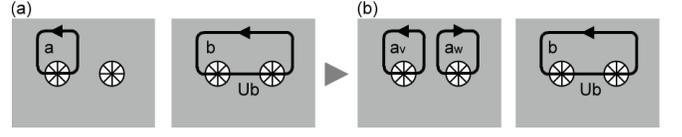}
\caption{(a) The generators $W^a_{\alpha_1}$ and $W^b_{\beta_1}$ for the group of physical Wilson loop operators on the Klein bottle. (b) By the structural property of the subgroup $\SS$, the modified generators $W^{a}_*=W^{a_v}_{\alpha_1}\ast W^{-a_w}_{\alpha_2}$ and $W^b_{\beta_1}$ can be treated as the raising and measuring operators.}\label{fig7}
\end{figure}

Now we construct the ground states on $\mathbb{K}$. Similar to the previous calculation, we figure out the group $\mathcal{W}(\mathbb{K})$ and get a generating set where the measuring and raising operators can be identified. Any physical Wilson operator along a loop in $H_1(\mathbb{K})$ can carry a symmetric anyon, so the generating set has $W^a_{\alpha_1}$ and $W^a_{\alpha_2}$ where $a\in \SS$. Since any physical Wilson operator along a two-sided loop can carry arbitrary anyon, the generating set also contains $W^b_{2\alpha_1}$, $W^b_{2\alpha_2}$ and $W^b_{\beta_1}$ where $b\in \AA$. The operators $W^b_{2\alpha_1}$ and $W^b_{2\alpha_2}$ carry anyon $b$ counter-clockwisely around the first and the second cross caps respectively.
The operator $W^b_{\beta_1}$ carries the anyon $b$ counter-clockwisely
from the second cross cap to the first cross cap and it carries $Ub$ from the first cross cap back to the second cross cap. The complete generating set for $\mathcal{W}(\mathbb{K})$ is $\{W^a_{\alpha_1},~W^a_{\alpha_2},~W^b_{2\alpha_1},~W^b_{2\alpha_2},~W^b_{\beta_1}\}$. Note that $W^a_{\alpha_2}=W^{-a}_{\alpha_1}\ast W^{a}_{\beta_1}$, $W^b_{2\alpha_i}=W^{(I+U)b}_{\alpha_i}$ where $i=1,~2$ and $(I+U)b\in\SS$. The generators $W^a_{\alpha_2}$, $W^b_{2\alpha_1}$ and $W^b_{2\alpha_2}$ can be expressed in terms of $W^a_{\alpha_1}$ and $W^b_{\beta_1}$, hence they can be eliminated from the generating set of $\mathcal{W}(\mathbb{K})$. Therefore,
\begin{align}\label{WK}
\mathcal{W}(\mathbb{K})=\langle~W^a_{\alpha_1},~W^b_{\beta_1}~|~a\in\SS,~b\in\AA~\rangle.
\end{align}
Such generating set is not good enough since $W^a_{\alpha_1}$ are not mutually commuting though $W^b_{\beta_1}$ are. Recall that
for any $a\in\SS$, we have $a=a_v+a_w$ for some $a_v\in V$ and $a_w\in W$. Denote $W^{a}_*=W^{a_v}_{\alpha_1}\ast W^{-a_w}_{\alpha_2}$. Since $W^{a}_{\alpha_1} \ast W^{-a_w}_{\beta_1}=W^{a}_*$, the generators $W^a_{\alpha_1}$ can be replaced by the operator $W^{a}_*$,
\begin{align}
\mathcal{W}(\mathbb{K})=\langle~W^{a}_*,~W^b_{\beta_1}~|~a_v\in V,~a_w\in W,~b\in\AA~\rangle.
\end{align}
Recall that the anyons in $V$ and $W$ are mutually bosonic saparately, so $W^{a}_*$ by such construction are mutually commuting. Hence we can simultaneously diagonalize $W^b_{\beta_1}$ and treat $W^{a}_*$ as the raising operators (Fig.\ \ref{fig7}) in constructing ground states.

%To construct the ground states on the Klein bottle $\mathbb{K}$, we treat $W^b_{\beta_1}$ as the measuring operators and  $W^{a_v}_{\alpha_1}\ast W^{-a_w}_{\alpha_2}$ as the raising operators.
The ground states corresponds to different fluxes measured by $W^b_{\beta_1}$. Let $e$ be the standard basis for $\mathbb{Z}^{N}$. The Wilson algebra on the Klein bottle $\mathbb{K}$ is given by
%\small
\begin{align}
W^{e_I}_{\beta_1}W^{a}_*
=e^{i2\pi e_I^TK^{-1}a}W^{a}_*W^{e_I}_{\beta_1}.
%\nonumber
\end{align}
%\normalsize
The measurement of flux vectors $e^TK^{-1}a$ defines a group homomorphism $f_{e}|_\SS:\SS \rightarrow \mathbb{Q}^{N}/\mathbb{Z}^{N};a\mapsto eK^{-1}a$. Consider a reference state $|0\rangle$ with $W^{e_I}_{\beta_1}|0\rangle=|0\rangle$.
A state $|a\rangle$ is constructed by operating the raising operator $W^{a}_*$ on the state $|0\rangle$.
Two states $|a\rangle$ and $|a'\rangle$ are the same if they get the same flux vector. So we have
\begin{align}\label{GSK}
&|a\rangle=W^{a}_*|0\rangle
~~\mbox{and}~~\\
&W^{e_I}_{\beta_1}|a\rangle=e^{i2\pi e_I^TK^{-1}a}|a\rangle,
\end{align}
where the anyon $a\in \SS/\textrm{Ker} f_e|_{\SS}\simeq f_e|_{\SS}[\SS]$. The quantity GSD$(\mathbb{K})$ corresponds to the number of flux vectors in the image $f_e|_{\SS}[\SS]=e^TK^{-1}u\mathbb{Z}^{M}/\mathbb{Z}^{N}$. Since $e$ is the standard basis for $\mathbb{Z}^{N}$, $e$ can be considered as the identity matrix of $N$ dimension. Hence we get $f_e|_{\SS}[\SS]\simeq \SS$. Therefore,
\begin{align}\label{GSDK}
\textrm{GSD}(\mathbb{K})=|\SS|,
\end{align}
which is the total number of parity symmetric anyons. Different ground states corresponds to different fluxes measured by the loop operator $W^b_{\beta_1}$ under the flux insertion by $W^{a}_*$.

%In the first homology group, we have the group relation $2(\alpha_1+\alpha_2)=0$. It requires the Wilson loop operator $W^{a}_{2(\alpha_1+\alpha_2)}$ to be trivial, and hence $W^{2a}_{\alpha_1+\alpha_2}$ must commutes with any physical Wilson operator. Since $u_i^{T}K^{-1}u_j=0~\textrm{or}~\frac{1}{2}~\textrm{mod}~1$, it commutes with the generator $W^a_{\alpha_1}$. Note that it trivially commutates with the other generator $W^b_{\beta_1}$. Hence it commutes with any physical Wilson operator.

%\begin{figure*}[t]
%\centering
%\includegraphics[scale=0.3]{DiffKleinBottles.eps}
%\caption{(a) The construction of a Klein bottle by a connected sum of two cross caps. (b) The construction of a Klein bottle by gluing the M\"obius strip. In the figures shown above, any parity twist is decoupled from the defect branch cut which is indicated by a dotted line decorated by a little arrow indicating the direction of the twist. The two constructions of Klein bottle are equivalent iff. the direction of the little arrows can be flipped, that is $\MM^2=I$.}\label{DiffK}
%\end{figure*}

\begin{figure*}[t]
\centering
\includegraphics[scale=0.26]{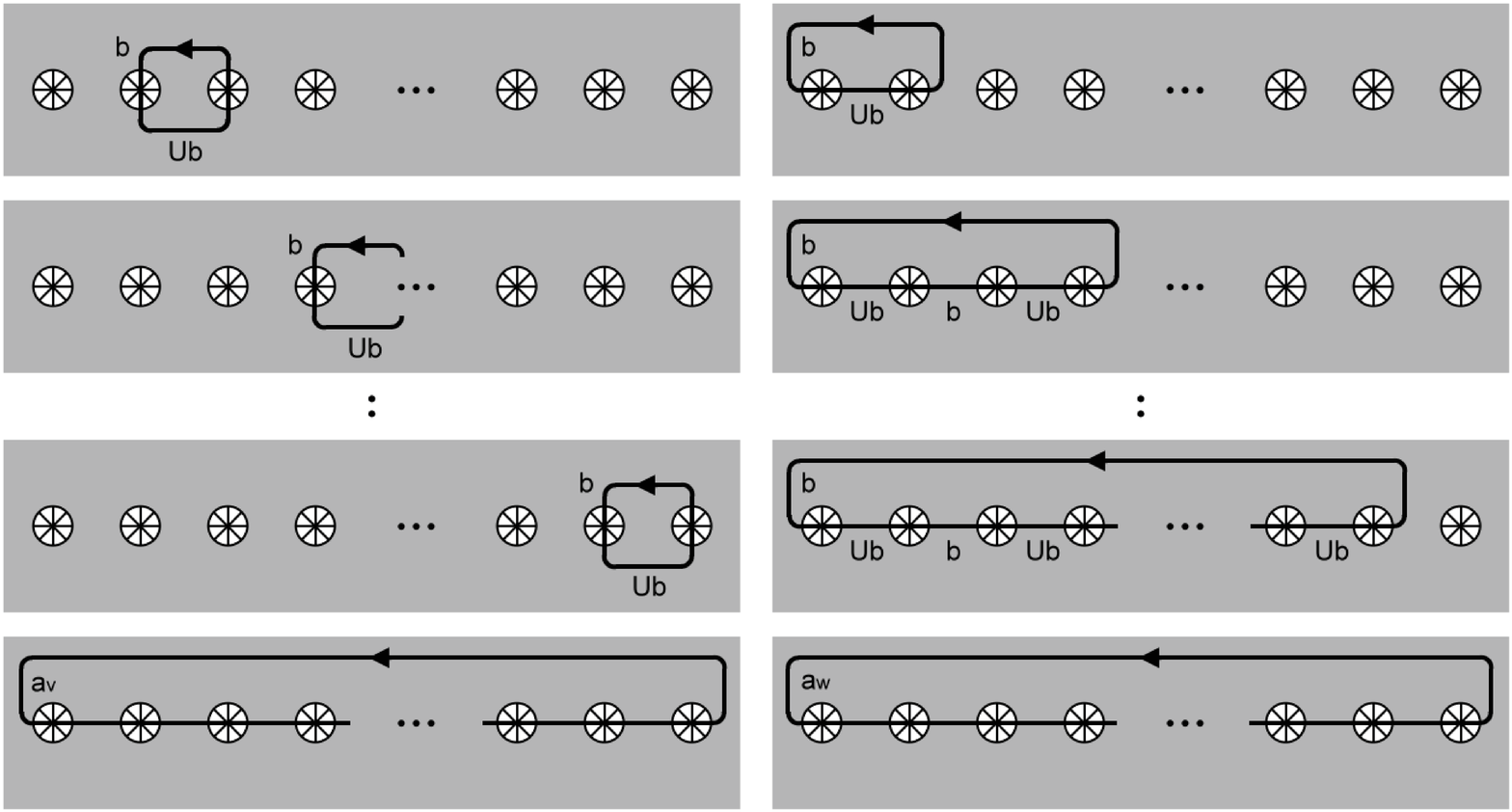}
\caption{Generators for the group of physical Wilson loop operators on $q\mathbb{P}$ where $q$ is odd. The generators in the first and the second column respectively correspond to the generators in the first and the second line in Eq.\ (\ref{genodd}). The figures on the first column show the raising operators and the figures on the second column show the corresponding measuring loop operators.}\label{fig8}
\includegraphics[scale=0.26]{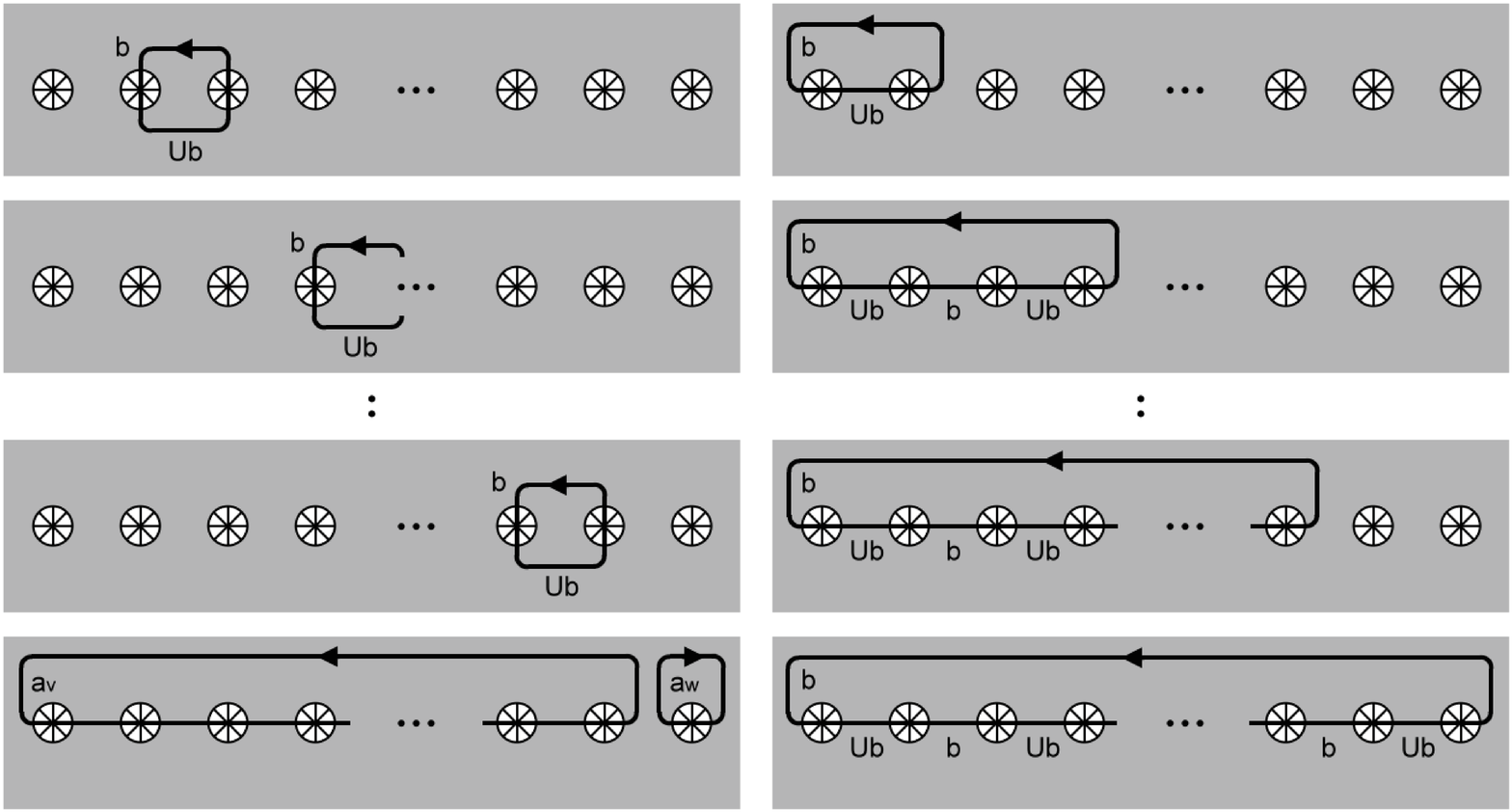}
\caption{Generators for the group of physical Wilson loop operators on $q\mathbb{P}$ where $q$ is even. The generators in the first and the second column respectively correspond to the generators in the first and the second line in Eq.\ (\ref{geneven}). The figures on the first column show the raising operators and the figures on the second column show the corresponding measuring loop operators.}\label{fig9}
\end{figure*}

\subsection{General Non-orientable Closed Surfaces}

The construction of states on real projective plane allow us to discuss states on general non-orientable closed manifolds. Mathematically, the set of closed non-orientable manifolds up to homeomorphism forms a commutative semigroup under the connected sum $\#$. The semigroup is generated by the real projective plane  $\mathbb{P}$,
\begin{align}
\{\textrm{Non-orientable Closed Surfaces}\}=\langle~\mathbb{P}~\rangle,
\end{align}
where connected sum with $\mathbb{P}$ is the same as adding a cross cap.
The surface $\mathbb{P}\#\mathbb{P}$ is the Klein bottle $\mathbb{K}$ whereas the surface $\mathbb{P}\#\mathbb{P}\#\mathbb{P}$ is the Dyck's surface.
With a state on real projective plane $\mathbb{P}$, we can generate a state on any non-orientable closed surface.
Let $q\mathbb{P}$ be the connected sum of $q\geq 1$ cross caps and let $\alpha_i$ be the one-sided loop associated with the $i$-th cross cap.
Its first homology is
\begin{align}
H_1(q\mathbb{P})
=\langle~\alpha_1,\dots,~\alpha_{q}~|~2\alpha_{T}=0~\rangle=\mathbb{Z}^{q-1}\times \mathbb{Z}_2,
\end{align}
where $\alpha_T=\alpha_1+\dots+\alpha_{q}$.
%In the first homology group, any loop which enters cross cap by odd number of times is one-sided whereas any loop which enters cross cap by even number of times is two-sided. The subgroup of two-sided loops in $H_1(\mathbb{K})$ is generated by the loops $2\alpha_1,\dots,~2\alpha_q$ and $\beta_1,\dots,~\beta_{q-1}$ where the loop $\beta_i$ is defined as $\beta_i=\alpha_{i}+\alpha_{i+1}$.
%The intersection number between any two loops in the first homology group can be obtained by $I(\alpha_i,~\alpha_{j})=\delta_{i,j}~\textrm{mod}~2$.
%Similar to the previous calculations, the group of physical Wilson operators can be obtained.
%Any Wilson operator along any loop in the first homology group can carry a symmetric anyon. In addition, any Wilson operator along any two-sided loop in the first homology group can carry arbitrary anyon. So the generating set for the group $\mathcal{W}(\mathbb{P}\#\dots\#\mathbb{P})$ is $\{W^a_{\alpha_1},\dots,~W^a_{\alpha_q},~W^b_{2\alpha_1},\dots,~W^b_{2\alpha_q},~W^b_{\beta_1},\dots,~W^b_{\beta_{q-1}}\}$
%The operator $W^b_{2\alpha_i}$ carry $b$ counter-clockwisely around the $i$-th cross caps.
Let $\beta_i=\alpha_i+\alpha_{i+1}$ and $W^b_{\beta_i}$ be the operator carrying $b$ counter-clockwisely from the $i+1$-th cross cap to the $i$-th cross cap and $Ub$ from the $i$-th cross cap back to the $i+1$-th cross cap.
%Within the generating set, some of the generators are redundant
%since $W^a_{\alpha_{i+1}}=W^{-a}_{\alpha_i}\ast W^{a}_{\beta_i}$ and $W^b_{2\alpha_i}=W^{(I+U)b}_{\alpha_i}$
%where $i=1,\dots,~q$ and $(I+U)b\in\SS$.
%From the second relation, we can get rid of the generators $W^b_{2\alpha_1},\dots,~W^b_{2\alpha_q}$,
%whereas from the first relation, we can eliminate the generators $W^a_{\alpha_2},\dots,~W^a_{\alpha_q}$.
%Therefore,
Then
\begin{align}\label{generalgen}
\mathcal{W}(q\mathbb{P})
=&\langle~W^a_{\alpha_1},~W^b_{\beta_1},\dots,W^b_{\beta_{q-1}}~|~a\in\SS,~b\in\AA~\rangle.
\end{align}
Note that $2\alpha_T=0$ in $H_1(q\mathbb{P})$, and it requires $W^{a}_{2\alpha_T}$ to be trivial. Since the entries of $[2u^{T}K^{-1}u]$ are in $\mathbb{Z}_2$, $W^{2a}_{\alpha_T}$ commute with the generators $W^a_{\alpha_1}$. Note that $W^{2a}_{\alpha_T}$ trivially commute with $W^b_{\beta_i}$'s since there is no intersection between $\alpha_T$ and $\beta_i$'s. Hence $W^{2a}_{\alpha_T}$ commute with all Wilson operators and the group relation is guaranteed.

%Next we are going to identify the measuring and raising operators in the generating set so as to construct the ground states and calculate the ground state degeneracy for states defined on general non-orientable closed surfaces. It turns out that we need to consider the cases with odd and even number of cross caps separately.

\begin{widetext}
\subsubsection{Odd Number of Cross Caps}

The construction of ground states for surfaces with odd number of cross caps is a generalization of the construction for the surface with one cross cap.
Given a non-orientable closed surface with odd number of
cross caps, we have $W^{a}_{\alpha_T}=W^{a}_{\alpha_1}\ast W^a_{\beta_2}\ast W^a_{\beta_4}\ast\dots\ast W^a_{\beta_{q-1}}$. Hence
%\begin{align}
%\mathcal{W}(q\mathbb{P})
%=
%\left\langle\begin{array}{ccccc}
%W^b_{\beta_2},&W^b_{\beta_4},&\dots,&W^b_{\beta_{q-1}},&W^{a}_{\alpha_T},\\
%W^b_{\beta_1},&W^b_{\beta_3},&\dots,&W^b_{\beta_{q-2}}&\\
%\end{array}
%\right.
%\left|\begin{array}{ccccc}
%a\in \SS,\\
%b\in\AA\\
%\end{array}
%\right\rangle,
%\end{align}
we can replace the generator $W^a_{\alpha_1}$ in Eq.\ (\ref{generalgen}) by the operator $W^{a}_{\alpha_T}$.
Further rewriting the generating set so as to identify the measuring and the raising operators, we get
\begin{align}\label{genodd}
\mathcal{W}(q\mathbb{P})
=
\left\langle\begin{array}{ccccc}
W^b_{\beta_2},&W^b_{\beta_4},&\dots,&W^b_{\beta_{q-1}},&W^{a_v}_{\alpha_T},\\
W^b_{\beta_1},&W^b_{\beta_1+\beta_3},&\dots,&W^b_{\beta_1+\dots+\beta_{q-2}},&W^{a_w}_{\alpha_T}\\
\end{array}
\right.
\left|\begin{array}{ccccc}
a_v\in V,~a_w\in W,\\
b\in\AA\\
\end{array}
\right\rangle.
\end{align}
Notice that the generators in the first line commute and the generators in the second line also commute (see Fig.\ \ref{fig8}). So we can simultaneously diagonalize the later generators and treat the former generators as the raising operators. Remarkably, non-trivial Wilson algebra appears only in each column and hence the Wilson algebras are decoupled.
% and are given by
%\begin{align}
%W^{e_I}_{\beta_1+\dots+\beta_{2p-1}}W^b_{\beta_{2p}}&=e^{i2\pi e_I^TK^{-1}b} W^b_{\beta_{2p}}W^{e_I}_{\beta_1+\dots+\beta_{2p-1}},\\
%W^{w_i}_{\alpha_T}W^{a_v}_{\alpha_T}&=e^{i2\pi w_i^TK^{-1}a_v} W^{a_v}_{\alpha_T}W^{w_i}_{\alpha_T},\label{odd al}
%\end{align}
%where $p=1,~2,\dots,~(q-1)/2$.
Each of the first $(q-1)/2$ algebras give $|\textrm{det}K|$ states while the last algebra gives $\textrm{GSD}(\mathbb{P})$ states. So for $q$ is odd,
\begin{align}\label{GSDoddP}
\textrm{GSD}(q\mathbb{P})=|\textrm{det}K|^{(q-1)/2}|\hat{\SS}|^{1/2}.
\end{align}
The ground states correspond to different fluxes distinguished by the measuring operators shown in second column of Fig.\ \ref{fig8}. Take $q=1$, the result above reduces to that of the real projective plane $\mathbb{P}$.

\subsubsection{Even Number of Cross Caps}

The GSD on manifolds with even number of cross caps is a generalization of the GSD on the surface with two cross caps, which is the Klein bottle. Consider a surface with even number of cross caps. In this case, we have $W^{a}_{\alpha_T-\alpha_{q}}=W^{a}_{\alpha_1}\ast W^a_{\beta_2}\ast W^a_{\beta_4}\ast\dots\ast W^a_{\beta_{q-2}}$ instead. So
%\begin{align}
%\mathcal{W}(q\mathbb{P})
%=
%\left\langle\begin{array}{ccccc}
%W^b_{\beta_2},&W^b_{\beta_4},&\dots,&W^b_{\beta_{q-2}},&W^{a}_{\alpha_T-\alpha_q},\\
%W^b_{\beta_1},&W^b_{\beta_3},&\dots,&W^b_{\beta_{q-3}},&W^b_{\alpha_{q-1}}\\
%\end{array}
%\right.
%\left|\begin{array}{ccccc}
%a\in \SS,\\
%b\in\AA\\
%\end{array}
%\right\rangle,
%\end{align}
we can replace the operator $W^{a}_{\alpha_1}$ in Eq.\ (\ref{generalgen}) by the operator $W^{a}_{\alpha_T-\alpha_{q}}$. To figure out the measuring and the raising operators, we further rewrite the generating set as
\begin{align}\label{geneven}
\mathcal{W}(q\mathbb{P})
=
\left\langle\begin{array}{ccccc}
W^b_{\beta_2},&W^b_{\beta_4},&\dots,&W^b_{\beta_{q-2}},&W^{a}_*,\\
W^b_{\beta_1},&W^b_{\beta_1+\beta_3},&\dots,&W^b_{\beta_1+\dots+\beta_{q-3}},&W^{b}_{\alpha_T}\\
\end{array}
\right.
\left|\begin{array}{ccccc}
a_v\in V,~a_w\in W,\\
b\in\AA\\
\end{array}
\right\rangle.
\end{align}
where $W^{a}_*=W^{a_v}_{\alpha_T-\alpha_q}\ast W^{-a_w}_{\alpha_q}$.
Note that the generators in the first line commute and the generators in the second line also commute (see Fig.\ \ref{fig9}). So we can simultaneous diagonalize the generators in the first line and treat the generators in the second line as the raising operators. Note that the Wilson algebras are decoupled into columns,
%\begin{align}
%W^{e_I}_{\beta_1+\dots+\beta_{2p-1}}W^b_{\beta_{2p}}&=e^{i2\pi e_I^TK^{-1}b} W^b_{\beta_{2p}}W^{e_I}_{\beta_1+\dots+\beta_{2p-1}},\\
%W^{e_I}_{\alpha_T}(W^{a_v}_{\alpha_T-\alpha_q}\ast W^{-a_w}_{\alpha_q})&=e^{i2\pi e_I^TK^{-1}a} (W^{a_v}_{\alpha_T-\alpha_q}\ast W^{-a_w}_{\alpha_q})W^{e_I}_{\alpha_T},
%\end{align}
%where $p=1,~2,\dots,~(q-2)/2$.
where each of the first $(q-2)/2$ algebras give $|\textrm{det}K|$ states and the last algebra gives $\textrm{GSD}(\mathbb{K})$ states. So for $q$ is even, we have
\begin{align}\label{GSDevenP}
\textrm{GSD}(q\mathbb{P})=|\textrm{det}K|^{(q-2)/2}|\SS|.
\end{align}
Each ground state is labeled by the fluxes identified by the measuring operators shown in the second column of  Fig.\ \ref{fig9}. If $q=2$, the GSD reduces to that of the Klein bottle $\mathbb{K}$.

%\begin{align}
%|\textrm{det}K|^{1/2}
%&=
%\left|\textrm{det}
%\scriptsize
%\left(\begin{array}{cccc}
%0&A&B&B\\
%A^T&0&C&-C\\
%B^T&C^T&E&D\\
%B^T&-C^T&D^T&-E\\
%\end{array}
%\right)\right|^{1/2}
%\normalsize
%=
%\left|\textrm{det}
%\scriptsize
%\left(\begin{array}{cccc}
%0&A&B&B\\
%A^T&0&-C&C\\
%B^T&C^T&D&E\\
%B^T&-C^T&-E&D^T\\
%\end{array}
%\right)\right|^{1/2}
%\normalsize\\
%&=
%\left|\textrm{det}
%\scriptsize
%\left(\begin{array}{cccc}
%0&A&B&B\\
%-A^T&0&C&-C\\
%-B^T&C^T&D&E\\
%-B^T&-C^T&-E&D^T\\
%\end{array}
%\right)\right|^{1/2}
%\normalsize
%=
%\left|\textrm{pf}
%\scriptsize
%\left(\begin{array}{cccc}
%0&A&B&B\\
%-A^T&0&C&-C\\
%-B^T&C^T&D&E\\
%-B^T&-C^T&-E&D^T\\
%\end{array}
%\right)\right|
%\normalsize
%\end{align}
\end{widetext}

From the results in Eqs.\ (\ref{GSDoddP}) and (\ref{GSDevenP}), we can extract the quantum dimension of a cross cap by taking the large $q$ limit. Since adding two cross caps generally gives rise to an extra $|\textrm{det}K|$ amount of independent ground states, the quantum dimension of a cross cap is simply given by $|\textrm{det}K|^{1/2}$, which must be an integer and is exactly the same as total quantum dimension $\mathcal{D}$ of the state $K$.

\subsection{Robustness of the Ground State Degeneracy}

Here we are going to show the robustness of the GSD
by checking the consistency of our result with the geometric Dyck's theorem.
The set of closed surfaces up to homeomorphism forms a commutative monoid under connected sum $\#$.
The identity element is the sphere,
and the monoid is generated by torus $\mathbb{T}$ and real projective plane $\mathbb{P}$ with a single relation, that is
\begin{align}
\{\textrm{Closed Surfaces}\}=\langle~\mathbb{T},~\mathbb{P}~|~\mathbb{P}\#\mathbb{P}\#\mathbb{P}=\mathbb{P}\#\mathbb{T}~\rangle.
\end{align}
The relation $\mathbb{P}\#\mathbb{K}=\mathbb{P}\#\mathbb{T}$ is known as the Dyck's theorem.
Geometrically,
$\#\mathbb{T}$ add a handle to the surface with the two ends attached on the same side
while $\#\mathbb{K}$ add a handle to the surface with the two ends attached on the opposite sides.
In the presence a real projective plane $\mathbb{P}$,
the surface is non-orientable, so there is no difference between $\#\mathbb{T}$ and $\#\mathbb{K}$.
Let $\Sigma(g,q)$ be a closed surface given by a connected sum of $g$ tori and $q$ cross caps.
The Dyck's theorem is equivalent to the statement that
\begin{align}
\Sigma(g,q+2)
=\Sigma(g+1,q),
\end{align}
for any $q\geq1$.
So the non-orientable surfaces $\Sigma(g,q+2)$ and $\Sigma(g+1,q)$ are homeomorphic.
We are going to prove that their GSD are the same so as to
check the consistency of our results with the Dyck's theorem and show the robustness of the GSD at the same time.

To show the consistency, we make use of the following observation.
The generating set of $H_1(\Sigma(g,q))$ composes of the generators of the $g$ tori and the generators of the $q$ cross caps. Let $\rho_j$ and $\sigma_j$ be the standard generators for the $j$-th handle and $\alpha_i$ be the generator for the $i$-th cross caps with $\beta_i=\alpha_i+\alpha_{i+1}$. The generating set for the group of physical Wilson operators on $\Sigma(g,q)$ are generated by the union of $\{~W^b_{\rho_1},~W^b_{\sigma_1},\dots,~W^b_{\rho_g},~W^b_{\sigma_g}~\}$ and $\{~W^a_{\alpha_1},~W^b_{\beta_1},\dots,~W^b_{\beta_{q-1}}~\}$ where $a\in\SS$, $b\in\AA$. Notice that former generators for the $g$ tori do not intersect with the later generators for the $q$ cross caps, so the Wilson algebra for the $g$ tori and the $q$ cross caps decouples. So
\begin{align}
\textrm{GSD}(\Sigma(g,q))
=\textrm{GSD}(\Sigma(g,0))\textrm{GSD}(\Sigma(0,q)),
\end{align}
where it is well known that the factor $\textrm{GSD}(\Sigma(g,0))=|\textrm{det}K|^{g}$.
By making use of the GSD we obtained for general non-orientable closed surfaces,
we can evaluate the other factor $\textrm{GSD}(\Sigma(0,q))$.
Consequently, we have the following consistency equation
\begin{align}
\textrm{GSD}(\Sigma(g,q+2))
%=&\textrm{GSD}(\Sigma(g,0))\textrm{GSD}(\Sigma(0,q+2))
%\nonumber \\
=&\textrm{GSD}(\Sigma(g,0))\textrm{GSD}(\Sigma(0,q))|\textrm{det}K|
\nonumber \\
%=&\textrm{GSD}(\Sigma(g+1,0))\textrm{GSD}(\Sigma(0,q))
%\nonumber \\
=&\textrm{GSD}(\Sigma(g+1,q)),
\end{align}
for any $q\geq1$. It shows that the GSD we obtained previously is consistent with the Dyck's theorem, i.e., GSD is a homomorphism from the monoid $\{\textrm{Closed Surfaces}\}$ to the monoid of positive integers.
Physically, the GSD is robust against deformation of two cross caps to a genus on a non-orientable surface.

\section{The mapping class group of the Klein Bottle}\label{S6}

In this section, we are going to discuss the physical meaning of the set of large diffeomorphisms, i.e., the mapping class group (MCG), of the non-orientable manifolds. Notice that the real projective plane $\mathbb{P}$ has trivial MCG, we focus on the first non-trivial case which is the MCG of the Klein bottle. First, we introduce the MCG on the Klein bottle and its group action on the group of physical Wilson loop operators. Then we proceed to its group action on the multiplet of ground states. Subsequently, we obtain a matrix representation and hence a physical interpretation for the MCG.

\subsection{The Mapping Class Group}

\begin{figure*}[t]
\centering
\includegraphics[scale=0.27]{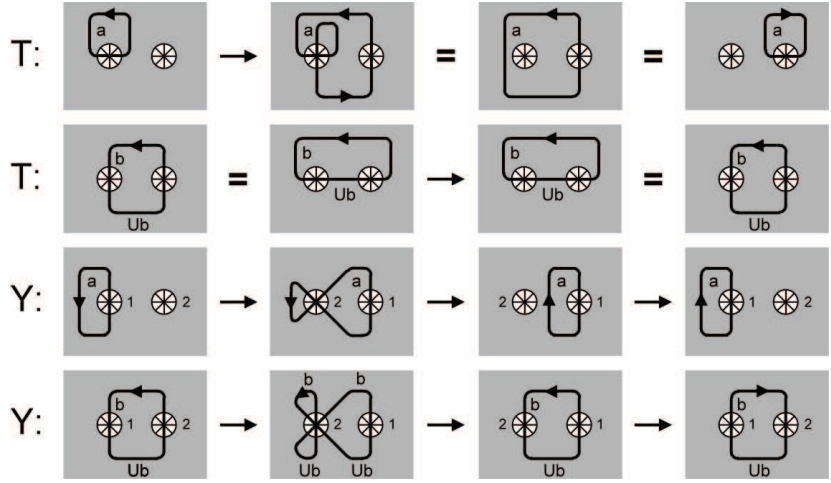}~~~~~~
\caption{The first two rows show respectively the action of the Dahn twist T on $W^{a}_{\alpha_1}$ and $W^{b}_{\beta_1}$.
An right arrow between two figures indicates the action under T.
An equal sign between two figures means a deformation of the loop.
The second two rows show respectively the action on $W^{a}_{\alpha_1}$ and $W^{b}_{\beta_1}$ under the Y-homeomorphism Y.
To be precise, the action under Y is split into three steps and the two cross caps are labeled by $1$ and $2$.
In the first two steps, the cross cap $1$ is passed through the cross cap $2$ from left to right. The third step is to drag the cross cap $1$ back to its original position.
\label{fig10}}
\end{figure*}

Here we introduce the MCG of the Klein bottle and talk about how its group elements act on the physical Wilson loop operators in $\mathcal{W}(\mathbb{K})$.
Let Aut$(X)$ be the group of automorphisms of a closed surface $X$.
Denote Aut$_0(X)$ as the subgroup of automorphisms that are isotopic to the identity.
The MCG of $X$ is defined as
\begin{align}
\textrm{MCG}(X)=\textrm{Aut}(X)/\textrm{Aut}_0(X),
\end{align}
which is the group of isotopy classes of automorphisms of $X$.
As proven by Dehn and rediscovered by Lickorish,
the MCG of any orientable closed surface $X$ is generated by Dehn twists \cite{DEHN,LICKORISH1}.
In addition,
Lickorish has shown that the MCG of any non-orientable closed surface $X$ with $q>1$ is generated by Dehn twists and one Y-homeomorphism (or the cross cap slide) \cite{LICKORISH2,LICKORISH3}. In particular,
\begin{align}
\textrm{MCG}(\mathbb{K})&=\langle~\textrm{T},~\textrm{Y}~|~\textrm{T}\textrm{Y}=\textrm{Y}\textrm{T},~\textrm{T}^2=\textrm{Y}^2=I~\rangle
%&=\mathbb{Z}_2\times\mathbb{Z}_2,
\end{align}
where the generator T is the Dehn twist and the generator Y is the Y-homeomorphism.
Under the Dehn twist T, the Klein bottle is twisted along the two-sided loop
$\beta_1$.
%[Fig.\ (\ref{Auto})].
On the other hand, under the Y-homeomorphism Y,
the first cross cap passes through the second cross cap and get back to its original position.
Essentially, the first cross cap has gone through a cycle of a M\"obius strip in a round trip.
%[Fig.\ (\ref{Auto})].

Note that the Dehn twist T and the Y-homeomorphism Y are automorphism for the geometric Klein bottle.
However, it is not necessarily true for a Klein bottle constructed by parity defects.
Notice that the configuration of parity branch cuts on the Klein bottle is always invariant under T.
It can be shown that the arrangement of parity branch cuts on the Klein bottle is invariant under Y iff. $\MM^2=I$.

%There is also restriction on $\chi$.
%Consider a Klein bottle $\mathbb{K}$ constructed by a parity symmetry with $\MM^2=I$.

Since the symmetry group $\mathbf{}\textrm{MCG}(\mathbb{K})$ acts on the surface $\mathbb{K}$,
it induces a group action on the group $\mathcal{W}(\mathbb{K})$ of physical Wilson loop operators.
As shown in Eq.\ (\ref{WK}), we can take $W^{a}_{\alpha_1}$ and $W^{b}_{\beta_1}$ as the generators for $\mathcal{W}(\mathbb{K})$.
From Fig.\ \ref{fig10},
we see that the actions on the generators under T and Y transformations,
\begin{align}
\textrm{T}W^a_{\alpha_1}\textrm{T}^{-1}&=W^{-a}_{\alpha_2},~~~~
\textrm{T}W^b_{\beta_1}\textrm{T}^{-1}=W^b_{\beta_1}.\label{TAction}\\
\textrm{Y}W^a_{\alpha_1}\textrm{Y}^{-1}&=W^{-a}_{\alpha_1},~~~~
\textrm{Y}W^b_{\beta_1}\textrm{Y}^{-1}
%=W^{Ub}_{\beta_1}
=W^{-b}_{\beta_1}.\label{YAction}
\end{align}
Since the association of an automorphism in $\mathbf{}\textrm{MCG}(\mathbb{K})$ with a group action on $\mathcal{W}(\mathbb{K})$ is a homomorphism. The group relation in $\mathbf{}\textrm{MCG}(\mathbb{K})$ are trivially satisfied by the induced group action.

%\begin{align}
%\textrm{Y}=\mathcal{C}~;~\textrm{T}=\mathcal{P}
%\end{align}
%\begin{figure*}[t]
%\centering
%\includegraphics[scale=0.3]{Auto.eps}~~~~~~
%\caption{The group action of $\textrm{MCG}(\mathbb{K})$ on the Klein bottle constructed by twisting a parity  symmetry $\MM$.
%In the figures shown above,
%\textcolor{blue}{any parity twist is decoupled from the branch cut which is indicated by a dotted line decorated by a little arrow indicating the direction of the twist.
%(Not sure what this sentence means...)}
%Such Klein bottle is trivially invariant under the Dehn twist T. It is invariant under the Y-homeomorphism iff. the direction of the little arrows can be flipped, that is $\MM^2=I$.
%\label{Auto}
%}
%\end{figure*}

\subsection{The Matrix Representation}

In the previous section, we have seen that elements in
$\textrm{MCG}(\mathbb{K})$ induce a group action on the group $\mathcal{W}(\mathbb{K})$ of physical Wilson loop operators. Consequently, it defines a group action operating on the ground states. In this section, we are going to calculate the matrix element of the generators T and Y for $\mathcal{W}(\mathbb{K})$.

Recall that the ground states on the Klein bottle are defined in Eq.\ (\ref{GSK}) where we have measuring operators $W^b_{\beta_1}$ and raising operators $W^{a}_*$. From Eq.\ (\ref{TAction}), we have $[\textrm{T},W^b_{\beta_1}]=0$, so we can simultaneously diagonalize the operators T and $W^b_{\beta_1}$. Also, by using Eqs.\ (\ref{TAction}), it can be shown that $\textrm{T}W^{a}_*\textrm{T}^{-1}=e^{i\theta_a}W^{a}_*W^{a}_{\beta_1}$,
%\begin{align}
%\textrm{T}W^{a_v}_{\alpha_1} W^{-a_w}_{\alpha_2}\textrm{T}^{-1}
%&=\textrm{T}W^{a_v}_{\alpha_1}\textrm{T}^{-1}\textrm{T}W^{-a_w}_{\alpha_2}\textrm{T}^{-1}\nonumber\\
%&=W^{-a_v}_{\alpha_2} W^{a_w}_{\alpha_1}\nonumber\\
%&=W^{a_w}_{\alpha_1} W^{-a_v}_{\alpha_2}\nonumber\\
%&=W^{a_v+(a_w-a_v)}_{\alpha_1} W^{-a_w+(a_w-a_v)}_{\alpha_2}\nonumber\\
%&=(e^{-i\pi a_v^TK^{-1}a_w}W^{a_v}_{\alpha_1}W^{a_w-a_v}_{\alpha_1})
%(e^{-i\pi a_w^TK^{-1}a_v}W^{-a_w}_{\alpha_2}W^{a_w-a_v}_{\alpha_2})\nonumber\\
%&=e^{-2i\pi a_v^TK^{-1}a_w}W^{a_v}_{\alpha_1}W^{-a_w}_{\alpha_2}W^{a_w-a_v}_{\beta_1}\nonumber\\
%&=e^{-i\theta_a}W^{a_v}_{\alpha_1}W^{-a_w}_{\alpha_2}W^{a}_{\beta_1}\nonumber\\
%&=e^{i\theta_a}W^{a_v}_{\alpha_1}W^{-a_w}_{\alpha_2}W^{a}_{\beta_1}
%\end{align}
%\begin{align}
%\textrm{T}W^{a}_*\textrm{T}^{-1}=e^{i\theta_a}W^{a}_*W^{a}_{\beta_1},
%\end{align}
which tells how the Dehn twist operates on the raising operators for the ground states. Therefore, we can immediately deduces the matrix element for the Dehn twist T,
%\begin{align}
%\langle a'|\textrm{T}|a\rangle
%&=\langle a'|\textrm{T}W^{a_v}_{\alpha_1} W^{-a_w}_{\alpha_2}|0\rangle\nonumber\\
%&=\langle a'|\textrm{T}W^{a_v}_{\alpha_1} W^{-a_w}_{\alpha_2}\textrm{T}^{-1}\textrm{T}|0\rangle\nonumber\\
%&=\langle a'|e^{i\theta_a}W^{a_v}_{\alpha_1}W^{-a_w}_{\alpha_2}W^{a}_{\beta_1}\textrm{T}|0\rangle\nonumber\\
%&=e^{i\theta_{\textrm{T}}}e^{i\theta_a}
%\langle a'|W^{a_v}_{\alpha_1}W^{-a_w}_{\alpha_2}W^{a}_{\beta_1}|0\rangle\nonumber\\
%&=e^{i\theta_{\textrm{T}}}e^{i\theta_a}
%\langle a'|W^{a_v}_{\alpha_1}W^{-a_w}_{\alpha_2}|0\rangle\nonumber\\
%&=e^{i\theta_{\textrm{T}}}e^{i\theta_a}\langle a'|a\rangle\nonumber\\
%&=e^{i\theta_{\textrm{T}}}e^{i\theta_a}\delta_{aa'}
%\end{align}
\begin{align}\label{MRT}
\langle a'|\textrm{T}|a\rangle
%&=\langle a'|\textrm{T}W^{a}_*\textrm{T}^{-1}\textrm{T}|0\rangle
%=\langle a'|e^{i\theta_a}W^{a}_*W^{a}_{\beta_1}\textrm{T}|0\rangle
%\nonumber \\
&=e^{i\theta_{\textrm{T}}}e^{i\theta_a}\delta_{aa'},
\end{align}
where $\textrm{T}|0\rangle=e^{i\theta_{\textrm{T}}}|0\rangle$.
Hence the matrix representation of Dehn twist T is diagonal with the diagonal entries tell the topological spins of the parity symmetric anyons.
%Since $\textrm{T}^2=I$, the parity  symmetric anyons are either self-bosons or fermion, which agrees with the result we obtained in Eq.\ (\ref{statpro}) by considering the restriction of $S$ and $T$ matrices to the subgroup $\SS$.

Similarly, we can obtain a matrix representation for the Y-homeomorphism Y.
Note from Eq.\ (\ref{YAction}) that Y flips the sign of the anyon label on any Wilson operator. Therefore, the Y-homeomorphism Y can be thought of as the charge conjugation operator operating on the ground states.
Hence we get the matrix elements
\begin{align}\label{MRY}
\langle a'|\textrm{Y}|a\rangle=e^{i\theta_{\textrm{Y}}}\delta_{\bar{a}a'},
\end{align}
where $\textrm{Y}|0\rangle=e^{i\theta_{\textrm{Y}}}|0\rangle$. Hence the matrix representation of the Y-homeomorphism Y tells the particle conjugation for the parity symmetric anyons.

%In the matrix representation of $\mathbf{}\textrm{MCG}(\mathbb{K})$, we have the group relation $\textrm{T}\textrm{Y}=\textrm{Y}\textrm{T}$ trivially satisfied. In the analysis here, we can only tell that $\textrm{T}^2=e^{i2\theta_{\textrm{T}}}$ and $\textrm{Y}^2=e^{i2\theta_{\textrm{Y}}}$ for the remaining group relations.

\section{Examples}\label{S7}

In this section, we are going to consider two examples of abelian states with $\mathbb{Z}_2$ parity symmetry mentioned in Appendix\ \ref{S4}. The first one is the fermionic state $K_f$ with $D=0$ and the second one is the bosonic state $K_b$. First, we obtain explicitly the generating set $u=v\cup w$ for the subgroup $\SS$ of parity symmetric anyons. Then we count the number of parity symmetric anyons and the number of distinguishable parity symmetric anyons using Eqs.\ (\ref{symsubgroup}) and (\ref{dissymsubgroup}) respectively. Next, we calculate the ground state degeneracies on arbitrary non-orientable closed surfaces. Finally, we write down the matrix representation of the generators for the mapping class group on the Klein bottle $\mathbb{K}$.

\subsection{The Fermionic State}

In this subsection, we consider the fermionic state $K_f$ with $D=0$ and we denote such fermionic state as $\tilde{K}_f$. Generally, the state can possess more than one $\mathbb{Z}_2$ parity symmetry, and we consider a particular one here. Let $E$ be a $n\times n$ invertible symmetric integer value matrix. The K-matrix $K=\tilde{K}_f$ and the matrix $U$ for the $\mathbb{Z}_2$ parity symmetry are respectively given by
\begin{align}\label{FS}
%\footnotesize
\tilde{K}_f=
\left(\begin{array}{cc}
E&0\\
0&-E\\
\end{array}
\right)
~~~~\mbox{and}~~~~
U=
\left(\begin{array}{cc}
0&~I~\\
~I~&0\\
\end{array}
\right),
%\normalsize
\end{align}
where the eigenvectors of $U$ with eigenvalue $+1$ are given by $v_i=(e_i~e_i)^T$ where $i=1,~2,\dots,~n$.
Take any $a=(a_1~a_2)^T\in\SS$, we have $Ua=a+K\Lambda$ for some integer vector $\Lambda=(\Lambda_1~\Lambda_2)^T$.
Such relation can be translated as $a_2=a_1+E\Lambda_1$  with $\Lambda_2=\Lambda_1$. Hence, up to a local particle, any anyon $a\in\SS$ can be written as $a=a_{1i}v_i~~\textrm{mod}~K\Lambda$,
%\begin{align}
%\footnotesize
%a
%=
%\left(\begin{array}{c}
%a_1\\
%a_1+E\Lambda_1\\
%\end{array}
%\right)
%&=
%\left(\begin{array}{c}
%a_1\\
%a_1\\
%\end{array}
%\right)
%+
%\tilde{K}_f
%\left(\begin{array}{c}
%0\\
%-\Lambda_1\\
%\end{array}
%\right)
%\simeq
%\left(\begin{array}{c}
%a_1\\
%a_1\\
%\end{array}
%\right)
%=a_{1i}v_i~~\textrm{mod}~K\Lambda,
%\normalsize
%\end{align}
for some $a_1\in\mathbb{Z}^n$. From the equation above, we can see that any parity symmetric anyon can be written as a linear combination of the eigenvectors $v_i$'s. Hence $\SS$ is generated by the set $u=v\cup w$ with
\begin{align}
v=\{v_1,\dots,~v_n\}
~~~~\textrm{and}~~~~
w=\emptyset,
\end{align}
where $\emptyset$ denotes an empty set. With the generating set $u$, we get the structure of the subgroup $\SS$ of parity symmetric anyons. By Eq.\ (\ref{symsubgroup}), one can show that $\SS\simeq E^{-1}\mathbb{Z}^{n}/\mathbb{Z}^{n}$.
%\begin{align}
%\SS
%&=u\mathbb{Z}^M/\tilde{K}_f\mathbb{Z}^N
%\simeq \tilde{K}_f^{-1}u\mathbb{Z}^{M}/\mathbb{Z}^{N}
%=\tilde{K}_f^{-1}v\mathbb{Z}^{n}/\mathbb{Z}^{N}
%\nonumber \\
%&=\left(\begin{array}{c}
%E^{-1}\\
%-E^{-1}\\
%\end{array}
%\right)
%\mathbb{Z}^{n}/\mathbb{Z}^{N}
%\simeq E^{-1}\mathbb{Z}^{n}/\mathbb{Z}^{n}.
%\end{align}
The total number of parity symmetric anyons is given by the number of lattice points, that is $|\SS|=|\textrm{det} E|$.
%\begin{align}
%|\SS|=|\textrm{det} E|.
%\end{align}
Observe that the generators in $u$ are mutually bosonic, the matrix $[2u^TK^{-1}u]$ vanishes. By Eq.\ (\ref{dissymsubgroup}), we get $|\hat{\SS}|=1$.
%\begin{align}
%|\hat{\SS}|=1,
%\end{align}
meaning that all the parity symmetric anyons are statistically indistinguishable in $\SS$
\footnote{From Appendix\ \ref{S4}, we know that the matrix $[2u^TK^{-1}u]$ generally vanishes for any fermionic state $K_f$, hence the result that $|\hat{\SS}|=1$ can be applied to any general fermionic state $K_f$.}.
Since the symmetric anyons in the state $\tilde{K}_f$ are mutually bosonic, all the physical Wilson operators on the real projective plane $\mathbb{P}$ shown in Fig.\ \ref{fig5} are mutually commuting. Since the Wilson algebra is trivial, there is a unique ground state on $\mathbb{P}$. Alternatively, there is no measuring operator $W^{a_w}_\alpha$ to distinguish different fluxes, hence
\begin{align}
\textrm{GSD}(\mathbb{P})=1
\end{align}
which is as expected from Eq.\ (\ref{GSDP}). On the Klein bottle $\mathbb{K}$, the measuring operators are $W^b_{\beta_1}$ whereas the raising operators are $W^{a_v}_{\alpha_1}$ as shown in Fig.\ \ref{fig7}. The number of different possible fluxes measured by $W^b_{\beta_1}$ is given by the number of flux insertion operators $W^{a_v}_{\alpha_1}$ which equals $|V|=|\SS|$. Therefore, we have the result
\begin{align}
\textrm{GSD}(\mathbb{K})
=|\textrm{det} E|.
\end{align}
which agrees with Eq.\ (\ref{GSDK}). Generally, adding two cross caps on an unorientable manifold gives extra $|\textrm{det}K|=|\textrm{det}E|^2$ ground states.
Hence the ground state degeneracy on a manifold with $q$ cross caps is given by
\begin{align}
\textrm{GSD}(q\mathbb{P})=|\textrm{det}E|^{q-1}.
\end{align}
The result above suggests that the state $\tilde{K}_f$ defined on an unorientable manifold is equivalent to the state $E$ defined on its orientable double cover
\footnote{The orientable double cover of $\Sigma(0,q)$ is $\Sigma(q-1,0)$.}. Beside, we can consider the MCG of $\mathbb{K}$. Since all the parity symmetric anyons are self-bosons, we have the result
\begin{align}
\langle a'|\textrm{T}|a\rangle
=e^{i\theta_{\textrm{T}}}\delta_{aa'}
~,~~
\langle a'|\textrm{Y}|a\rangle=e^{i\theta_{\textrm{Y}}}\delta_{\bar{a}a'},
\end{align}
which gives the matrix representation of the generators of the mapping class group on the Klein bottle $\mathbb{K}$.

%\begin{figure}[t]
%\centering
%\includegraphics[scale=0.3]{Kleinon.eps}
%\caption{
%(a) A pair of defects for the on-site layer permutation symmetry $\tau$ in the bilayer quantum hall state $K_{bi}$ can be viewed as a genon bridging the two layers $E$. (b) Flipping the lower layer in $K_{bi}$, we get the state $\tilde{K}_f$. Meanwhile the genon is deformed into the Kleinon bridging the upper layer $E$ and the flipped layer $-E$, where its handle penetrates through the lower layer at some place. Such Kleinon is equivalent to a pair of parity defects or a cross cap in $\tilde{K}_f$}\label{fig11}
%\end{figure}

To understand the parity defects we constructed for the state $\tilde{K}_f=E\oplus-E$, we consider a more intuitive construction from the genon \cite{TWISTSDEF} in a bilayer quantum hall state $K_{bi}=E\oplus E$ with an on-site $\mathbb{Z}_2$ layer permutation symmetry $\tau$. The transformation matrix for the on-site symmetry $\tau$ in the state $K_{bi}$ is exactly the same as the matrix $U$ for the parity symmetry of the state $\tilde{K}_{f}$. Consider a pair of defects for $\tau$ connected by an on-site branch cut placed along the $x$-axis and centered at origin in the state $K_{bi}$. Such pair of defects for the layer permutation symmetry can be viewed as a genon
%bridging the bottom face of the upper layer $E$ and the top face of the lower layer $E$ in the state $K_{bi}$
\cite{TWISTSDEF}.
%(Fig.\ (\ref{fig11}a))
Considering flipping over the lower layer about the axis $x=0$, the bilayer quantum hall state $K_{bi}$ becomes the fermionic state $\tilde{K}_f$. It can be checked that the on-site gapping term for $\tau$ turns into the gapping term for parity symmetry by sending $x$ to $-x$ for the lower layer edge variables in Eq.\ (\ref{gapping}). Therefore, the resulting pair of parity defects can be viewed geometrically as a deformed genon.
%bridging the bottom face of the upper layer $E$ and the bottom face of the lower layer $-E$ in the state $\tilde{K}_{f}$, where its handle penetrates through the lower layer at some place (Fig.\ (\ref{fig11}b)). We call such deformed genon the Kleinon.
Note that the anyon label of any anyon in a layer going though the genon and the deformed genon is always unchanged. However any oriented loop going through the genon preserves its orientation whereas any oriented loop going through the
%Kleinon
deformed genon flips its orientation. It can also be checked that the GSD contributed by a
%Kleinon
deformed genon in $\tilde{K}_{f}$ is the same as that contributed by a genon in $K_{bi}$. In fact, both of them are of quantum dimension $|\textrm{det} E|$.

To end this example for the fermionic state $\tilde{K}_f$, we consider the particular case where $E$ is an integer with $E=2$. In such case, the state $\tilde{K}_f$ corresponds to the double semion state. Remarkably, in such state, the $\mathbb{Z}_2$ parity symmetry is unique and is shown in Eq.\ (\ref{FS}). In the double semion state, we have the group of anyons $\AA=\{1,~s_1,~ s_2,~\xi\}$ where $1$ is the vacuum; $s_1$ and $s_2$ are mutually bosonic semions; $\xi$ is a boson formed by the fusion of $s_1$ and $s_2$. The parity symmetry relabels the anyons as $(1,~s_1,~ s_2,~\xi)\mapsto(1,~s_2,~ s_1,~\xi)$. Hence the subgroup of parity symmetric anyons is $\SS=\{1,~\xi\}$ which is generated by $u=v=\{\xi\}$. Since all the parity symmetric anyons in $\SS$ have trivial mutual-statistics, they cannot be distinguish from one another in $\SS$ and hence $\hat{\SS}=\{[1]\}$ in which $\xi$ get identified with the vacuum $1$. Consequently, we have $\textrm{GSD}(\mathbb{P})=1$ and $\textrm{GSD}(\mathbb{K})=2$. Generally, on the connected sum of $q$ cross caps, we have $\textrm{GSD}(q\mathbb{P})=2^{q-1}$.

\subsection{The Bosonic State}

In this subsection, we consider the bosonic state $K_b$. Let $A$ be a $n\times n$ non-singular integer value matrix. Such topological state may possess more than one $\mathbb{Z}_2$ parity symmetry and we pick a particular one. The K-matrix $K=K_b$ and the transformation matrix $U$ for the $\mathbb{Z}_2$ parity symmetry are respectively given by
\begin{align}\label{BS}
%\footnotesize
K_b=
\left(\begin{array}{cc}
0&A\\
A^T&0\\
\end{array}
\right)
~~~~\mbox{and}~~~~
U=
\left(\begin{array}{cc}
-I&0\\
0&I\\
\end{array}
\right),
\normalsize
\end{align}
where the eigenvectors of $U$ with eigenvalue $+1$ are given by $v_i=(0~e_i)^T$ where $i=1,~2,\dots,~n$.
For any $a=(a_1~a_2)^T\in\SS$, we have $Ua=a+K\Lambda$ for some integer vector $\Lambda=(\Lambda_1~\Lambda_2)^T$. Such relation can be written as $a_1=\frac{1}{2}A(-\Lambda_2)$ with $\Lambda_1=0$. Hence $\forall a\in\SS$, we have $a=a_{2i}v_i+(\frac{1}{2}A\Lambda_2~0)^T~\textrm{mod}~K\Lambda.$
%\begin{align}\label{symab}
%a
%&=
%\left(\begin{array}{c}
%\frac{1}{2}A(-\Lambda_2)\\
%a_2\\
%\end{array}
%\right)
%\nonumber \\
%&=
%\left(\begin{array}{c}
%\frac{1}{2}A\Lambda_2\\
%a_2\\
%\end{array}
%\right)
%+
%K
%\left(\begin{array}{c}
%0\\
%-\Lambda_2\\
%\end{array}
%\right)
%\simeq
%\left(\begin{array}{c}
%\frac{1}{2}A\Lambda_2\\
%a_2\\
%\end{array}
%\right).
%=a_{2i}v_i+
%\left(\begin{array}{c}
%\frac{1}{2}A\Lambda_2\\
%0\\
%\end{array}
%\right)~~\textrm{mod}~K\Lambda.
%\end{align}
Each integer vector from $\frac{1}{2}A\Lambda_2$ can give rise to a parity symmetric anyon. In other words, each $\Lambda_2\in\mathbb{Z}^{n}$ with $[A\Lambda_2]=0$ can lead to a parity symmetric anyon. Let $([~]\circ A)\Lambda_2=[A\Lambda_2]$ and $\lambda=\{\lambda_1,~\lambda_2,\dots,~\lambda_{n'}\}$ be a generating set of $\textrm{Ker}([~]\circ A)$. We have the generating set $u=v\cup w$ with
\begin{align}
v=\{v_1,\dots,~v_n\}
~~~~\textrm{and}~~~~
w=\{w_1,\dots,~w_{n'}\},
\end{align}
where $w_i=(\frac{1}{2}A\lambda_i~0)^T$.
With the generating set $u$, we can obtain the structure for the subgroup $\SS$. By Eq.\ (\ref{symsubgroup}), one can show that $\SS\simeq A^{-1}\mathbb{Z}^{n}/\mathbb{Z}^{n}\times[\lambda]\mathbb{Z}^{n'}_2$.
%\begin{align}
%\SS
%&=u\mathbb{Z}^M/K\mathbb{Z}^N
%\simeq K^{-1}u\mathbb{Z}^{M}/\mathbb{Z}^{N}
%=K^{-1}(v\cup w)\mathbb{Z}^{M}/\mathbb{Z}^{N}
%\nonumber \\
%&=\left(\begin{array}{cc}
%A^{T-1}&0\\
%0&\frac{1}{2}\lambda\\
%\end{array}
%\right)
%\mathbb{Z}^{M}/\mathbb{Z}^{N}
%\footnotesize
%\simeq A^{T-1}\mathbb{Z}^{n}/\mathbb{Z}^{n}\times[\lambda]\mathbb{Z}^{n'}_2.
%\end{align}
The number of lattice points in such lattice gives the total number of parity symmetric anyons, hence we have $|\SS|=|\textrm{det} A| 2^{\textrm{Rk}[\lambda]}$.
%\begin{align}
%|\SS|=|\textrm{det} A| 2^{\textrm{Rk}[\lambda]}.
%\end{align}
In the generating set $u=v\cup w$, it can be checked that the integer vectors in $v$ are mutually bosonic and the integer vectors in $w$ are also mutually bosonic. However, integer vectors in $v$ can braids non-trivially with integer vectors in $w$. More precisely, the matrix $[2w^T\tilde{K}_f^{-1}v]=[\lambda^T]$. By using Eq.\ (\ref{dissymsubgroup}), we have $|\hat{\SS}|=2^{2\textrm{Rk}[\lambda]}$.
%\begin{align}
%|\hat{\SS}|=2^{2\textrm{Rk}[\lambda]},
%\end{align}
Since $[~]\circ A=[A]\circ[~]$, it can be shown that $[\lambda]$ generates Ker$[A]$. Hence $\textrm{Rk}[\lambda]=\textrm{dim~Ker}[A]=n-\textrm{Rk}[A]$ where the last equal sign follows from the rank-nullity theorem. On the real projective plane $\mathbb{P}$, we treat $W^{a_w}_\alpha$ as the raising operators and $W^{a_v}_\alpha$ as the raising operators as shown in Fig.\ \ref{fig5}. The non-trivial Wilson algebra leads to a GSD determined by Eq.\ (\ref{GSDP}), that is
\begin{align}
\textrm{GSD}(\mathbb{P})
=2^{n-\textrm{Rk}[A]}.
\end{align}
On the Klein bottle $\mathbb{K}$, the measuring operator are $W^b_{\beta_1}$ whereas the raising operators are  $W^{a}_*$ as shown in Fig.\ \ref{fig7}. By using Eq.\ (\ref{GSDK}), we obtain the GSD
\begin{align}
\textrm{GSD}(\mathbb{K})
=|\textrm{det} A|2^{n-\textrm{Rk}[A]}.
\end{align}
Since adding two cross caps to an unorientable manifold contributes extra $|\textrm{det}K|=|\textrm{det}A|^2$ independent ground states. For surface constructed by a connect sum of $q$ cross caps, we have the general result
\begin{align}
\textrm{GSD}(q\mathbb{P})
=|\textrm{det} A|^{q-1}2^{n-\textrm{Rk}[A]}.
\end{align}
For example, consider the case $n=1$ where $A$ reduces to an non-zero integer and the bosonic state corresponds to a $\mathbb{Z}_A$ topological state. Finally, we consider a generic deformation of the Klein bottle $\mathbb{K}$. Notice that $e^{i\theta_a}=e^{i\pi a_2^T\Lambda_2}$. Hence we get the result
\begin{align}
\langle a'|\textrm{T}|a\rangle
=e^{i\theta_{\textrm{T}}}e^{i\pi a_2^T\Lambda_2}\delta_{aa'}
~,~~
\langle a'|\textrm{Y}|a\rangle=e^{i\theta_{\textrm{Y}}}\delta_{\bar{a}a'},
\end{align}
which gives the matrix representation of the Dehn twist
T and the Y-homeomorphism Y.

To end the discussion for the bosonic state, we consider the case where $A$ is an integer with $A=2$. The resulting state is a $\mathbb{Z}_2$ toric code where there is a unique $\mathbb{Z}_2$ parity symmetry and is given in Eq.\ (\ref{BS}). The group of anyons in the toric code is $\AA=\{1,~e,~ m,~\psi\}$ where $1$ is the vacuum; $e$ and $m$ are mutually semionic bosons; $\psi$ is a fermionic composite particle of $e$ and $m$. Under the parity symmetry, the anyons transform as $(1,~e,~ m,~\psi)\mapsto(1,~e,~ m,~\psi)$, meaning that all of the anyons are parity symmetric. The group $\SS=\{1,~e,~ m,~\psi\}$ is generated by $u=v\cup w$, where $v=\{m\}$ and $w=\{e\}$. Since $\SS=\AA$, we have $\tilde{\SS}=\{[1],~[e],~ [m],~[\psi]\}$ where there is no identification between the symmetric anyons.  Consequently, we have $\textrm{GSD}(\mathbb{P})=2$ and $\textrm{GSD}(\mathbb{K})=4$. Generally, on the connected sum of $q$ cross caps, $\textrm{GSD}(q\mathbb{P})=2^{q}$. Therefore, on any given non-orientable closed manifold, the GSD of the $\mathbb{Z}_2$ toric code is always double that of the double-semion state.

\section{Summary and Discussion}

In this paper, we explore the physical properties of topological states put on non-orientable surfaces. We introduce, for any given parity symmetry, the gapping term for the parity branch cut, which glues two edges of a state with a spatial flip. We construct a ribbon by gluing the opposite edges of a square-shaped topological state using the gapping term for parity. After the surgery, we find that carrying an oriented loop along the cycle of the ribbon flips the orientation of the loop, meaning that the ribbon is actually a M\"obius strip. On such twisted ribbon, only those anyons which are symmetric under the given parity symmetry are left. These parity symmetric anyons, which have their left hand statistics the same as their right hand statistics, determines the physical properties of the state when put on non-orientable surfaces constructed by suitable arrangement of the given parity branch cut. In other words, by studying the properties of the state on non-orientable surfaces, we can learn about the physical properties of these parity symmetric anyons. For example, the GSD on the real projective plane $\mathbb{P}$ equals $|\hat{\SS}|^{1/2}$, which is the root of the number of distinguishable parity symmetric anyons; the GSD on the Klein bottle $\mathbb{K}$ equals the total number $|\SS|$ of parity symmetric anyons. Generally, adding two cross caps on a non-orientable surface leads to an extra $|\textrm{det}K|$ independent ground states, which is exactly the number of independent ground states arise in adding a handle. Consequently, such GSD is robust against smooth deformation of two corss caps to a handle on top of a non-orientable surface. Besides, we study the action of the MCG on the ground states of the Klein bottle. We find that the Dehn twist T encodes the topological spins of the parity symmetric anyons whereas the Y-homeomorphism tells the particle-hole relation of the parity symmetric anyons. Finally, we work on the fermion state $\tilde{K}_f$ and the bosonic state $K_b$ as examples to illustrate our results.

While topological actions cannot be defined purely on non-orientable surfaces in a coordinate independent way, introducing parity branch cuts opens a window for us to study topological states on non-orientable surfaces. One particularly interesting scenario is the case where all anyons in $\AA$ are parity symmetric, e.g., the toric code model. In such case, the parity acts only on the anyon position without changing the anyon label. In addition, the twisted boundary condition for the bulk gauge field becomes continuous up to a physically trivial part
\footnote{If all anyons in $\AA$ are symmetric, we have $Ue_I=e_I+K\Lambda_I$ for some integer vector $\Lambda_I$. The twisted boundary condition for $a_x$ can be written as $e^T_Ia_{1x}(x)=e^T_I\tilde{a}_{2x}(sx)+\Lambda^T_I K\tilde{a}_{2x}(sx)$ where the last term gives trivial commutators quantum mechanically.}.
Since the state $K$ is modular, we have $\AA=\SS\simeq\hat{\SS}$. Hence the GSD on a connected sum of $q$ cross caps equals $|\textrm{det}K|^{q/2}$. The consequence of requiring the gauge field $a_\mu$ to be strictly continuous is discussed in a previous related work \cite{Chen2014}. In their work, they quantize the BF theory on non-orientable surfaces with the consideration of smooth differential forms and densities. Under such requirement, the GSD of toric code reduces from $4$ to $2$ on the Klein bottle $\mathbb{K}$. However, in the context of topological phases, even if all anyons in $\AA$ are parity symmetric, there is no reason in requiring $a_\mu$ to be strictly continuous across the branch cut.

In discussing the physical Wilson loop operators on the surfaces decorated with parity branch cuts, we require that any Wilson operator must have its finial anyon the same as its initial anyon and any Wilson loop is living in the first homology group of the surface. While the former condition guarantees that the Wilson loop operators do not create any gapped excitations, the later condition is imposed for Wilson loops on orientable surface \cite{CSQ, BFQ}. Generally on any surface, regardless of its orientability, the relevant loop group is the first homology group. To justify this, it suffices to show that the line integral of the gauge field along any homologically trivial loop is zero. Note that if a loop is homologically trivial, it is the boundary of some $2$-chain which is a formal sum of oriented $2$-simplexes. Since the flux through each of the $2$-simplex is zero, the flux through the $2$-chain is also zero. Hence, by Stokes' theorem, the line integral vanishes.

There are some open questions along the direction of this work. One open question is the derivation of the edge theory on non-orientable surfaces with non vanishing boundary, e.g., the M\"obius strip $\mathbb{M}$, which possesses a single edge. It would be interesting to know how the loss of orientation in the bulk affects the spectrum at the edge. Another interesting direction is the calculation of topological entanglement entropy and entanglement spectrum on a non-orientable subsystem, say, the M\"obius strip $\mathbb{M}$. It is known that the topological entanglement entropy on a disc shaped subsystem is given by $\ln \mathcal{D}$, which is independent on the choice of ground state \cite{TOPOENEN, BSTEE}. Also there is a one-to-one correspondence between the physical edge spectrum of the disc shaped subsystem and the low-lying spectrum of its entanglement hamiltonian \cite{ESNES1, ESNES2}. An immediate question is whether the same thing hold true on a M\"obius strip shaped subsystem, which shows up when non-orientable surfaces are suitably bipartited. Last but not least, a future direction is to generalize the discussion to put topological states on higher dimensional non-orientable spatial manifolds. By stacking up layers of $(2+1)$D topological states and suitably introducing coupling between layers, we can obtain $(3+1)$D topological states \cite{BTIs, LAYERCON}. By using similar techniques, we can stack up the parity defects to obtain a reflection defect in $(3+1)$D. Alternatively, one can construct such reflection defect by coupling adjacent layers with a parity twist. Such reflection defect can be used to define layers-constructed topological states on non-orientable spatial manifold.

\appendix
\begin{widetext}

\section{Abelian States with $\mathbb{Z}_2$ parity symmetry}\label{S4}

Here, we review the general data of the K-matrix theory for abelian topological states with a $\mathbb{Z}_2$ parity symmetry. Then we talk about a structural property of the subgroup $\SS$ of parity symmetric anyons. Mathematically, $\MM$ is a $\mathbb{Z}_2$ parity symmetry of a state $K$ if it is a symmetry of the state $K$ with $s=-1$ and $U^2=I$. In other words, $K$ is a state with a  $\mathbb{Z}_2$ parity symmetry if there exists a unimodular $U$ such that
\begin{align}\label{ParityZ2}
-UKU^T=K
~~~~\mbox{and}~~~~
U^2=I.
\end{align}
The equations above serves as the constraints for the matrices $K$ and $U$. If $K$ and $U$ is a solution to the above equations, then $K'$ and $U'$ is a gauge equivalent solution if there exists a unimodular $G$ such that $K'=GKG^{T}$ and $U'=GUG^{-1}$.
Up to gauge equivalence, the general solution \cite{MLSPT,CPTSYM} to Eq.\ (\ref{ParityZ2}) is given by
\begin{align}\label{ParityZ2Data}
K=
\left(\begin{array}{cccc}
0&A&B&B\\
A^T&0&C&-C\\
B^T&C^T&E&D\\
B^T&-C^T&D^T&-E\\
\end{array}
\right)
~~~~\mbox{and}~~~~
U=
\left(\begin{array}{cccc}
-I&0&0&0\\
0&~I~&0&0\\
0&0&0&~I~\\
0&0&~I~&0\\
\end{array}
\right)~,
\end{align}
where $A, B, C, D, E$ are integer value matrices with $D$ is anti-symmetric and $E$ is symmetric.
Note that $A$ is a $n_b\times n_b$ matrix whereas $B$ and $C$ are both $n_b\times n_f$ matrices.
Also matrices $D$ and $E$ are both of dimension $n_f\times n_f$. Hence Dim $K=N=2n$ where $n=n_b+n_f$. The state with $n_f=0$ is called the bosonic state $K_b$ whereas the state with $n_b=0$ is called the fermionic state $K_f$. Any local particle must be a boson in $K_b$ whereas the system may include local fermions in $K_f$.

We proceed to talk about the structure of the subgroup $\SS$ of parity symmetric anyons. More precisely, we are going to show that if a  state $K$ possesses a $\mathbb{Z}_2$ parity symmetry $\MM$, then there exists a set $v$ of $n=N/2$ mutually bosonic integer vectors and a set $w$ of $n'\geq0$ mutually bosonic integer vectors such that $u=v\cup w$ generates $\SS$. The set $v$ is precisely the integer eigenbasis of $U$ with eigenvalue $+1$. The set $u=v\cup w$ is called the generating set for $\SS$ induced by the $\mathbb{Z}_2$ parity symmetry.

First, we show that $u=v\cup w$ generates the subgroup $\SS$ of parity symmetric anyons for some set $w$ of integer vectors. Induced by the symmetry transformation $U$, we have a natural basis for the free abelian group $\mathbb{Z}^N$,
\begin{align}
v_{b~i_b}=
\left(\begin{array}{c}
0\\
e_{i_b}\\
0\\
0\\
\end{array}
\right)
~,~~
v_{f~i_f}=
\left(\begin{array}{c}
0\\
0\\
e_{i_f}\\
e_{i_f}\\
\end{array}
\right)
~,~~
x_{b~i_b}=
\left(\begin{array}{c}
e_{i_b}\\
0\\
0\\
0\\
\end{array}
\right)
~,~~
x_{f~i_b}=
\left(\begin{array}{c}
0\\
0\\
e_{i_f}\\
0\\
\end{array}
\right)
~,
\end{align}
where $i_{b/f}=1,~2,\dots,~n_{b/f}$.Let $v$ and $x$ be the collection of first two and last two species of basis vectors respectively, where we have $n$ basis vectors in both $v$ and $x$. By such construction, $v$ is the eigenbasis for $U$ with eigenvalue $+1$. Let $u_o$ be a generating set of the subgroup $\SS$.
Since $v$ is a subset of $\SS$, the union $v\cup u_o$ is also a generating set.
Note that we can expand any vector in $u_o$ in the basis $v\cup x$. Since $v$ is a subset of the generating set $v\cup u_o$, we can eliminate the $v$ components of the basis vectors in $u_o$ and denote the resultant set as $w$. Hence we obtain the generating set $u=v\cup w$ for $\SS$,
where any vector in $w$ can be expanded in the basis $x$, that is $w\subset x\mathbb{Z}^N$.
%By suitably removing the redundant vectors in $w$, we can further make the vectors in $w$ linearly independent.
We denote the number of vectors in such $w$ as $n'\geq0$. The subgroup generated by $v$ is denoted as $V=v\mathbb{Z}^n/K\mathbb{Z}^N$ whereas the subgroup generated by $w$ is denoted as $W=w\mathbb{Z}^{n'}/K\mathbb{Z}^N$.

Next, we show that symmetric anyons in $V$ are mutually bosonic and the anyons in $W$ are also mutually bosonic. For any $a\in\SS$, we have $Ua=a+K\Lambda$ for some $\Lambda\in\mathbb{Z}^N$. From the defining properties of $\mathbb{Z}_2$ parity symmetry in Eq.\ (\ref{ParityZ2}), we get $U^T\Lambda=\Lambda$.  Take any $a,~a'\in \SS$, since $U^T\Lambda=\Lambda$, we have $\Lambda^TK\Lambda'=0$. By expanding the expression $a^TU^TK^{-1}Ua'$, we get a formula for the statistics of symmetric anyons under the $\mathbb{Z}_2$ parity symmetry,
\begin{align}\label{Symanstat}
a^TK^{-1}a'=-\frac{1}{2}(a^T\Lambda'+a'^T\Lambda).
\end{align}
Since $U^T\Lambda=\Lambda$, $\Lambda$ is an eigenvector of $U^T$ with eigenvalue equals $+1$. Adopting the basis naturally induced by the parity symmetry as shown in Eq.\ (\ref{ParityZ2Data}), for any symmetric anyon $a$, we can write $a$ and $\Lambda$ in column vectors as
\begin{align}\label{Lambda}
a=
\left(\begin{array}{c}
a_{b1}\\
a_{b2}\\
a_{f1}\\
a_{f2}\\
\end{array}
\right)
,~~
\Lambda=
\left(\begin{array}{c}
0\\
\lambda_{b}\\
\lambda_{f}\\
\lambda_{f}\\
\end{array}
\right)
,
\end{align}
where the components $a_{b1},~a_{b2},~\lambda_{b}\in\mathbb{Z}^{n_b}$ and the components $a_{f1},~a_{f2},~\lambda_{f}\in\mathbb{Z}^{n_f}$.
In particular, if $a\in V$, we have $a_{b1}=0,~a_{f1}=a_{f2}$ and $\lambda_{b}=\lambda_{f}=0$.
Similarly, if $a\in W$, we have $a_{b2}=a_{f2}=0$.
By making use of the explicit form of $K$, $U$, $a$, $\Lambda$ in Eqs.\ (\ref{ParityZ2Data}) and (\ref{Lambda}), the defining equation $Ua=a+K\Lambda$ translates to
\begin{align}
a_{b_1}=-\frac{1}{2}(A\lambda_b+2B\lambda_f)
~~~~\mbox{and}~~~~
a_{f1}-a_{f2}=-C^T\lambda_b-(D+E)\lambda_f,
\end{align}
where $C\lambda_f=0$. Given the matrix $K$, the equations above relates the vectors $a$ and $\Lambda$.
Substitute the relations above into the formula in  Eq.\ (\ref{Symanstat}), we get the useful formula for the statistics of the parity symmetric anyons,
\begin{align}
a^TK^{-1}a'
%&=-\frac{1}{2}(a^T\Lambda'+a'^T\Lambda)\nonumber\\
%&=-\frac{1}{2}(a^T\Lambda'+\Lambda^Ta')\nonumber\\
%&=-\frac{1}{2}\Big((a_{b_2}^T\lambda'_{b}+\lambda^T_{b}a'_{b_2})
%+(a_{f_1}^T\lambda'_{f}+\lambda^T_{f}a'_{f_1})
%+(a_{f_2}^T\lambda'_{f}+\lambda^T_{f}a'_{f_2})\Big)\nonumber\\
%&=-\frac{1}{2}\Big((a_{b_2}^T\lambda'_{b}+\lambda^T_{b}a'_{b_2})
%+((a_{f_1}-a_{f_2})^T\lambda'_{f}+\lambda^T_{f}(a'_{f_1}-a'_{f_2}))
%+2(a_{f_2}^T\lambda'_{f}+\lambda^T_{f}a'_{f_2})\Big)\nonumber\\
%&=-\frac{1}{2}\Big((a_{b_2}^T\lambda'_{b}+\lambda^T_{b}a'_{b_2})
%-((\lambda_{b}^TC+\lambda_{f}^T(D+E)^T)\lambda'_{f}+\lambda^T_{f}(C^T\lambda'_{b}+(D+E)\lambda'_{f}))
%+2(a_{f_2}^T\lambda'_{f}+\lambda^T_{f}a'_{f_2})\Big)\nonumber\\
%&=-\frac{1}{2}\Big((a_{b_2}^T\lambda'_{b}+\lambda^T_{b}a'_{b_2})
%-2\lambda_{f}^TE\lambda'_{f}
%+2(a_{f_2}^T\lambda'_{f}+\lambda^T_{f}a'_{f_2})
%-\lambda_{b}^TC\lambda'_{f}-\lambda_{b}'^{T}C\lambda_{f}-\lambda_{f}^{T}(D+D^T)\lambda'_{f}\Big)\nonumber\\
&=-\frac{1}{2}(a_{b_2}^T\lambda'_{b}+a'^T_{b_2}\lambda_{b})
+\Big(\lambda_{f}^TE\lambda'_{f}
-(a_{f_2}^T\lambda'_{f}+a'^T_{f_2}\lambda_{f})\Big)
+\frac{1}{2}\Big(\lambda_{b}^TC\lambda'_{f}+\lambda_{b}'^{T}C\lambda_{f}+\lambda_{f}^{T}(D+D^T)\lambda'_{f}\Big)\nonumber\\
&~~~~~~~~~~~~~~~~~~~~=-\frac{1}{2}(a_{b_2}^T\lambda'_{b}+a'^T_{b_2}\lambda_{b})
+\Big(\lambda_{f}^TE\lambda'_{f}
-(a_{f_2}^T\lambda'_{f}+a'^T_{f_2}\lambda_{f})\Big),
\end{align}
where we made use of the facts that $C\lambda_f=C\lambda'_f=0$ and $D$ is anti-symmetric in the second equal sign. In the second line, the object inside the big round bracket is an integer which does not contribute to the mutual-statistics. Take any $a,a'\in V$, then $\lambda_{b}=\lambda'_{b}=\lambda_{f}=\lambda'_{f}=0$,
hence we have $a^TK^{-1}a'=0$.
Similarly, take any $a,a'\in W$, then $a_{b2}=a'_{b2}=0$, hence we have $a^TK^{-1}a'\in \mathbb{Z}$.
Therefore, the symmetric anyons in $V$ are mutually bosonic and the anyons in $W$ are also mutually bosonic.
\end{widetext}

\bibliography{biblio}
\bibliographystyle{phaip}
\pagestyle{plain}
\end{document}